\documentclass[10pt]{article}

\usepackage{amsmath}
\usepackage{amssymb}

\usepackage{graphicx}

\usepackage{cite}

\usepackage{color}

\usepackage[labelfont=bf,labelsep=period,justification=raggedright]{caption}

\makeatletter
\renewcommand{\@biblabel}[1]{\quad#1.}
\makeatother

\date{}

\usepackage{bussproofs}

\usepackage[all]{xy}
\usepackage{subcaption}
\usepackage{hyperref}
\hypersetup{colorlinks=true,
linkcolor=[rgb]{.67 .27 .27},
citecolor=[rgb]{.09 .29 .54},
urlcolor=[rgb]{.09 .29 .54}}

\usepackage[table]{xcolor}
\usepackage{rotating}
\usepackage{listings}
\usepackage{setspace}
\definecolor{Code}{rgb}{0,0,0}
\definecolor{Decorators}{rgb}{0.5,0.5,0.5}
\definecolor{Numbers}{rgb}{0.5,0,0}
\definecolor{MatchingBrackets}{rgb}{0.25,0.5,0.5}
\definecolor{Keywords}{rgb}{.67, .27, .27}
\definecolor{self}{rgb}{0,0,0}
\definecolor{Strings}{rgb}{.09,.29,.54}
\definecolor{Comments}{rgb}{0,0.5,0}
\definecolor{Backquotes}{rgb}{0,0,0}
\definecolor{Classname}{rgb}{0,0,0}
\definecolor{FunctionName}{rgb}{0,0,0}
\definecolor{Operators}{rgb}{0,0,0}
\definecolor{Background}{rgb}{0.98,0.98,0.98}

\lstnewenvironment{python}[1][]{
\lstset{
numbers=left,
numberstyle=\footnotesize,
numbersep=1em,
xleftmargin=1em,
framextopmargin=2em,
framexbottommargin=2em,
showspaces=false,
showtabs=false,
showstringspaces=false,
frame=l,
tabsize=4,
basicstyle=\ttfamily\small\setstretch{1},
backgroundcolor=\color{Background},
language=Python,
commentstyle=\color{Comments}\slshape,
stringstyle=\color{Strings},
morecomment=[s][\color{Strings}]{"""}{"""},
morecomment=[s][\color{Strings}]{'''}{'''},
morekeywords={import,from,class,def,for,while,if,is,in,elif,else,not,and,or,print,break,continue,return,True,False,None,access,as,,del,except,exec,finally,global,import,lambda,pass,print,raise,try,assert},
keywordstyle={\color{Keywords}\bfseries},
morekeywords={[2]@invariant},
keywordstyle={[2]\color{Decorators}\slshape},
emph={self},
emphstyle={\color{self}\slshape},
}}{}
 
\usepackage{placeins}

\definecolor{DeepRed}{rgb}{.82,.14,.16}
\definecolor{DeepBlue}{rgb}{0,0.36,0.62}

\renewcommand*{\sectionautorefname}{Sec.}

\newcommand{\beginsupplement}{        \setcounter{table}{0}
        \renewcommand{\thetable}{S\arabic{table}}        \setcounter{figure}{0}
        \renewcommand{\thefigure}{S\arabic{figure}}        \setcounter{equation}{0}
        \renewcommand{\theequation}{S\arabic{equation}}        \setcounter{section}{0}
        \renewcommand{\thesection}{S\arabic{section}}        \renewcommand*{\sectionautorefname}{Sec.}
        \setcounter{page}{1}
     }

\newcommand{\expr}{\mathcal{E}}
\newcommand{\dist}{\mathcal{D}}
\newcommand{\gnpm}{network-network state maps}
\newcommand{\refsupp}{Supplementary Material}
\newcommand{\AI}{AI}
\newcommand{\SH}{SH}

\begin{document}

\let\ref\autoref

\pagenumbering{arabic}

\begin{center}
{\large
\textbf{Potential unsatisfiability of cyclic constraints on stochastic biological networks biases selection toward hierarchical architectures}
}
\\[.5cm]
Cameron Smith$^{1}$,
Ximo Pechuan$^{1}$,
Raymond S. Puzio$^{1}$,
Daniel Biro$^{1}$,
Aviv Bergman$^{1,2,3,4, \ast}$
\\[.5cm]
$^1$Department of Systems and Computational Biology,\\
$^2$Dominick P. Purpura Department of Neuroscience,\\
$^3$Department of Pathology, Albert Einstein College of Medicine,\\
1301 Morris Park Ave, Bronx, NY 10461, USA\\
$^4$Santa Fe Institute, 1399 Hyde Park Road, Santa Fe, NM 87501, USA
\\[.5cm]
$\ast$To whom correspondence should be addressed; E-mail: aviv@einstein.yu.edu.
\end{center}

{ \bf

{\noindent\large Abstract}

Constraints placed upon the phenotypes of organisms result from their interactions with the environment.  Over evolutionary timescales, these constraints feed back onto smaller molecular subnetworks comprising the organism. The evolution of biological networks is studied by considering a network of a few nodes embedded in a larger context. Taking into account this fact that any network under study is actually embedded in a larger context, we define network architecture, not on the basis of physical interactions alone, but rather as a specification of the manner in which constraints are placed upon the states of its nodes. We show that such network architectures possessing cycles in their topology, in contrast to those that do not, may be subjected to unsatisfiable constraints. This may be a significant factor leading to selection biased against those network architectures where such inconsistent constraints are more likely to arise. We proceed to quantify the likelihood of inconsistency arising as a function of network architecture finding that, in the absence of sampling bias over the space of possible constraints and for a given network size, networks with a larger number of cycles are more likely to have unsatisfiable constraints placed upon them. Our results identify a constraint that, at least in isolation, would contribute to a bias in the evolutionary process toward more hierarchical-modular versus completely connected network architectures. Together, these results highlight the context-dependence of the functionality of biological networks.

}

\vskip .7em
\setcounter{secnumdepth}{4}

\section{Introduction}
Probabilistic models of biological networks serve as a bridge between theory and experiment.  On the one hand, parameters in a probabilistic model can be fit to data obtained by measuring the levels of each variable. For example in gene regulatory networks, gene expression can be measured using microarray or sequence census methods \cite{Anastassiou2007,Friedman2008a,Zhang2013}.  On the other hand, one can model a biological network as a deterministic or stochastic reaction network which tracks levels of each molecule \cite{Alon2006,Voit2012}.  From the solution to this latter kind of model, one can then obtain theoretical predictions for the parameters of the probabilistic model in terms of reaction rates.  Comparison of the parameters fitted from data with the predicted values serves as a means for comparing theory with experiment and can serve as a starting point for improving the theory or for designing future experiments \cite{Tonsing2014}.

An important feature of experimental science is that it involves partial information.  In the course of a single measurement, one typically is not able to observe a biological network in its entirety.  Rather, one observes a subnetwork at a time and only obtains a more complete picture by later combining these partial views.  This contrasts with theory, where, one makes a representation of a closed system that provides explicit values for all quantities of interest.  In order for a probabilistic model to serve its purpose, it should also accomodate partial information and thus we will explicitly consider the effects of 1) carving out a subnetwork from its context and 2) coarse-graining observables. Observables representing partial information will generally arise in situations where a system is interacting with another system. This situation arises in the context of interpreting the potential existence of modular substructure within biological network data deriving from any given organism as well as with respect to the interactions between an organism and its environment.

Inconsistency arises when a network context places more constraints on a subnetwork than it is capable of satisfying. The impact of this issue on genetic interactions has been considered previously in the context of population genetics \cite{EthanAkin389}. We exhibit a method of checking for such consistency and evaluating its likelihood of arising in the context of building probabilistic models of biological networks. When apparent inconsistency is observed, it must arise from the network context interacting with only partial information of the states of a given subnetwork. This would indicate that information about the network context must be included in order to maintain a consistent model of the system.

In \autoref{sec:networkcontext} we describe the relationship between representations of biological networks and an abstraction of these referred to as network architecture that indicates the manner in which a subset of a network is connected to its context. We explain the connection between stochastic process models of biological networks and a generalization of the genotype-phenotype map applying to arbitrary biological networks referred to as \gnpm{} in \autoref{sec:genenetworkphenmap}. \autoref{sec:probabilitydistributionsonnetworks}--\autoref{sec:inconsistency} contain examples of the underlying mathematical justification for our claims (more details of which are provided in \refsupp{}), and they can be skipped by readers who are primarily interested in the intuitive implications of our analysis. In \autoref{sec:probabilitydistributionsonnetworks} we introduce the concept of network modules and define probability distributions over their states. \autoref{sec:compatibilityofgpms} and \autoref{sec:inconsistency} describe the different compatibility conditions that arise for different biological network architectures and demonstrate how these compatibility conditions lead to a set of inequalities determining a space of probability distributions for each network architecture. \autoref{sec:cycliccontextunsatisfiableconstraints} and \autoref{sec:probconstrgeometry} examine these constraints for the example of the three-cycle network architecture. \autoref{sec:volrat} computes the likelihood of unsatisfiable constraints for all biological network architectures on four variables that possess cycles. Finally, \autoref{sec:unsatisfiableconstrevolution} explains implications for the evolution of biological network architectures of the result that networks with a larger number of cycles are more likely to have unsatisfiable constraints placed upon them.

\section{Environments of biological networks as abstract contexts}\label{sec:networkcontext}
\begin{figure}[!ht]
\centering
\noindent\includegraphics[width=0.9\columnwidth]{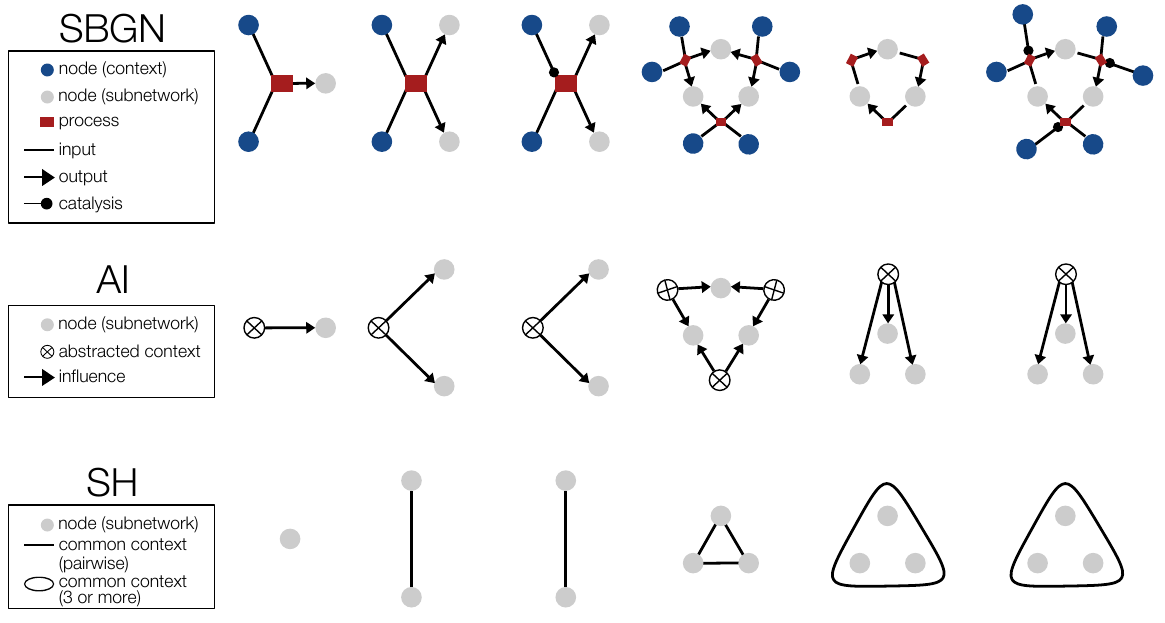}
\caption{{\bf Abstract influence representation of biological networks.} (row~1)~The systems biology graphical notation (SBGN) is capable of representing arbitrary biological networks including processes that involve metabolites, signaling molecules, genes, and enzymes \cite{LeNovere2009}. Only a fragment of the SBGN language, where all nodes have equivalent types, is indicated here. (row~2)~We abstract from the SBGN representation of a biological network to a graph representing the abstract influence (\AI{}) graph indicating coupling among a subset of the entities present in a biological network. (row~3)~For economy of representation we use a short hand (\SH{}) hypergraph to denote the \AI{} graph. The topology of the \AI{} and \SH{} graphs are equivalent and this is what we refer to as network architecture.}
\label{fig:netsubnetcontext}
\end{figure}

Most studies of biological networks focus on one type of variable in isolation. For example, many studies focus on one of metabolic networks, protein-protein interaction networks, signalling networks, gene-regulatory networks, or population and community dynamics in the context of ecological networks. A true biological network involves all of these acting together to produce biological phenomena at all scales. Models that integrate information about biological networks, rather than focusing exclusively on particular types of molecules, will likely become more common in the near future \cite{Covert2008,Karr2012,Macklin2014}. The Systems Biology Graphical Notation (SBGN) supports the ability to express many of these networks within the context of a single formalism \cite{LeNovere2009}, \autoref{fig:netsubnetcontext}. Even when the different types of biological variables are combined into a single network, it is impossible to study all variables simultaneously. As a result, it is always the case that a subnetwork is selected for investigation and the remainder of the network is treated as an \emph{environment} or \emph{context}. In \autoref{fig:netsubnetcontext} we show the SBGN process form of six simple examples of biological networks. In each case we have selected a subset of variables that form a subnetwork as an example of how one might proceed in the investigation of a particular biological system. Once such a subnetwork is chosen, it is possible to abstract away the variables that are not part of the subnetwork. This is represented by the abstract influence network (\AI{}) for each simple example on the second row of \autoref{fig:netsubnetcontext}. The transformation from SBGN to the \AI{} network is given simply by collapsing the disconnected components of the ancestors of each node in the focal subnetwork into single \AI{} nodes. This results in a bipartite graph that captures the dependencies among the environmental factors as experienced by the subnetwork and nothing more.

This \AI{} graph is precisely equivalent to an undirected hypergraph if one considers each of the \AI{} nodes as a hyperedge containing all of the nodes to which it connects. This is shown as the \SH{} graph in the third row of \autoref{fig:netsubnetcontext} for each of the simple examples of the SBGN form of biological networks. Considering all possible hypergraphs of this kind is equivalent to examining all possible environmental dependency structures the subnetwork could be subjected to. Because the \AI{} is fundamental to understanding how subnetworks depend upon their contexts, it is the structure of the \AI{} and equivalent \SH{} graphs that we refer to as \emph{network architecture} throughout the paper. We note from this perspective, that cycles in the SBGN representation of the biological network do not result in corresponding cycles in the \AI{} graph and vice versa. For instance, in example four of \autoref{fig:netsubnetcontext}, there are no cycles in the SBGN representation of the biological network whereas a single cycle exists in the hypergraph representation of the \AI{} graph. Furthermore, in example six, there is a cycle in the SBGN representation, whereas there is no cycle in the hypergraph representation of the \AI{}.

More precisely, the collection of variables comprising the subnetwork under consideration is referred to as $L$. The different subsets, $O$, of biological variables, $L$, making up the hypergraph representation of the \AI{} are each referred to as modules. A biological network architecture, $\mathcal{G}$, may then be represented by a subset of all possible such modules subject to two conditions (see \refsupp{} \autoref{sec:covergenotypespace}). The first represents the fact each variable of the focal subnetwork must be included in at least one module. The second represents the fact that any pair of constraints that are imposed upon overlapping sets of variables must agree on those overlapping variables.  In expressing the latter condition, all of the information present in a collection of lower-order constraints can be expressed as an effective higher-order constraint if any such higher-order constraint exists at all. So, if there is a constraint that is imposed simultaneously upon two distinct variables and another independent constraint imposed upon only the first of the two variables, this situation can be expressed in terms of a single constraint on both of the two variables.

When there is a relatively larger degree of independence in the network context as compared to the subnetwork, it is possible for inconsistency to arise. One canonical example of such inconsistency arises in the study of ferromagnetism via the Ising model on a triangular lattice where so-called \emph{frustration} arises in the couplings among the magnetic dipole moments of three nearest-neighbor atomic spins \cite{Wannier1950,Toulouse1977,Vannimenus1977}. In this example, the underlying lattice or graph represents interactions among the spins of atomic nuclei according to their spatial proximity. As we have described, in our model, the network architectures to which we refer represent the manner in which the network context places constraints upon a subnetwork. Inconsistency is likewise capable of arising if there is a cycle in the hypergraph representing this network architecture.

\section{Coarse-graining dynamic network states as a generalization of genotype-phenotype maps}\label{sec:genenetworkphenmap}

\autoref{fig:expression_concept}A shows a simplified representation of two different biological networks the correlation strengths among whose variables are not known but are to be derived from observation of the levels of the entities corresponding to each variable. For example, in the context of a gene-regulatory network, the amount of a given transcript present in a cell can be binned into a smaller number of discrete classes by setting a collection of thresholds on the original data set. If only a single threshold is given, then the data can be binned into two classes depending upon whether or not the original measurement surpasses the given threshold in \autoref{fig:expression_concept}B.
The time series that results from such observations can be used to infer various statistics that characterize the dynamics of a biological network such as correlations between pairs of variables.

If a large enough number of thresholds is available to distinguish among all possible counts of the variables under investigation, then this observational protocol becomes complementary to mechanistic models.  There may be several sources for stochasticity in the dynamics including small numbers of the causal molecules and products as well as environmental fluctuations upon which these dynamics are conditioned~\cite{Swain2002,Paulsson2004,Thattai2004,Acar2008a,Lestas2010,Munsky2012,Chalancon2012,Neuert2013,Sanchez2013}. Regardless of the fundamental nature of biological networks with respect to their potential stochasticity, empirical observations are usually regarded in a statistical manner, and thus we focus here on stochastic models.
Mathematically, such a model may take the form of a Markov chain whose dynamics are governed by a master equation for probability distributions over molecule counts. For example, in the case of a three variable network, the master equation takes the form
$$
\frac{dP(n_1,n_2,n_3)}{dt} = \sum_{n'_1}\sum_{n'_2}\sum_{n'_3} M^{n_1\,n_2\,n_3}_{n'_1\,n'_2\,n'_3}(k) P(n'_1,n'_2,n'_3)
$$
where $P(n_1,n_2,n_3)$ gives the probability of observing $n_1$, $n_2$, and $n_3$ molecules of each of the three variables respectively and $M(k)$ is a Markov transition rate matrix that depends upon some rate functions $k$ that are determined by the network architecture and the dynamics of the interactions.  The solution to this equation will converge towards a stationary distribution $P_s$ in the limit of long times. Any environmental variable having a characteristic timescale longer than that of the variables in the focal subnetwork would not be sensitive to transients and would only exhibit control over or be influenced by this stationary distribution.

Interactions between variables may be mediated by a coarse-graining over counts of each variable using a function that maps the states representing molecule counts as vectors of natural numbers into some other variables. For example, if $n_i$ are natural numbers
then a function $f$ taking any number less than or equal to some threshold $T$ to $0$ and any number greater than $T$ to $1$ is a very simple example of such a coarse-graining. For this specific form of the coarse-graining function $f$, the coarse-grained stationary probability distribution takes the form
$$
P_{cg}(b_1,b_2,b_3) = \sum_{n_1 \in f^{-1}(b_1)}\sum_{n_2 \in f^{-1}(b_2)}\sum_{n_3 \in f^{-1}(b_3)} P_{cg}(n_1,n_2,n_3),
$$
where $b_1,b_2,b_3 \in \{ 0,1 \}$. It is also possible to consider the case where each variable is coarse-grained according to a different threshold and into a different number of classes. An abstract algebraic formulation of the coarse-graining process is provided in \refsupp{} \autoref{secsupp:coarsegrainingphenotypes}.

The most familiar example of such a coarse-graining process in biology is the genotype-phenotype map. The genotype of an organism has a relatively straightforward definition in terms of the sequence of nucleotides comprising its genome. Phenotypes, on the other hand, can be described at different levels of organization~\cite{Dawkins1982,Stadler2001}. The concept of phenotype was initially defined at the level of macroscopically observable physical characteristics such as shape, size, color, and various combinations thereof~\cite{Johannsen1911}. However, since the advent of molecular biology, an example of a lower-level mapping upon which the higher-level map from molecular states to macroscopic phenotypes depends is the dynamic phenomenon that can be described by measuring the transcription states of all genes comprising an organism's genome.  These expression levels of subsets of interacting genes determine which enzymes are produced, thus determining the rate at which metabolic reactions proceed.  These reaction rates could then be viewed as constituting the next level of phenotypes.  These in turn determine even higher level phenotypes, ultimately culminating in macroscopically observable ones where the concept of phenotype was originally introduced. In summary, any mapping from the states of an underlying collection of molecules to a higher-level collective property of those molecules that may result from their interaction can be viewed as a generalization of the genotype-phenotype map, where the original conception of the latter corresponds to the special case where 1) the genes alone are sufficient to determe the higher-level collective property and 2) that higher-level collective property is observable at the whole-organism level.

A more realistic basis upon which to build phenotypes than this outline of the historical trajectory contains is one that is not limited to genes alone, but includes all entities constituting a biological network. A phenotype must be a function of the levels of, for example, all of the molecular constituents that comprise it over time, even if more information is required to fully specify it.  The aforementioned coarse-grained levels of biological network variables can thus be viewed as collectively determining the lowest level in a hierarchy of abstract phenotypes. In what proceeds, we will assume that we have a finite set $L$ of variables and a finite set $P$ of coarse-grained levels of each of those variables. These levels may have different units, but they can all be mapped into unitless quantities that account for the relevant scale of each variable. In general, each variable could take values in a distinct set $P_i, \; i \in I$ ranging over the variables, whereby $P$ would be required to represent $\cup_{i \in I} P_i$ rather than a monolithic valuation set lacking any underlying substucture with respect to the variables under consideration.
Then a possible state of our biological network is represented by a function $e : L \to P$ and coarse-graining a stationary distribution will lead to a probability distribution on the set of all maps, denoted $P^L$, from subnetworks represented by subsets of $L$ to the respective states of the variables that comprise them. We will refer to this more fine-grained generalization of the genotype-phenotype map, where arbitrary biological networks are substituted for genes and arbitrary networks states are substituted for phenotypes, as \gnpm{}.

\section{Probability distributions over network modules}\label{sec:probabilitydistributionsonnetworks}
Here we describe examples of probability distributions over network modules. A more general presentation is provided in \refsupp{} \autoref{secsupp:probabilitydistributionsonnetworks}.
As explained in \autoref{sec:networkcontext}, for a given biological subnetwork, the hypergraph representing the dependencies in the network context consists of subsets, $O$, of the variables, $L$, in the subnetwork.
If we consider the case in which we have two variables $L=\{l_1,l_2\}$ and there are two values, $P=\{0,1\}$, then there are four possible assignments of values to variables each of which constitutes a state of the system. We will write the probability of each of these states as $p^{v_1v_2}_{s_1s_2}$ indicating that variable $v_1$ is assigned value $s_1$ and variable $v_2$ is assigned value $s_2$. A probability distribution over the states of the system for $L$ is then given by
\begin{equation}\label{eq:examplejointdist}
\{p^{12}_{00},p^{12}_{01},p^{12}_{10},p^{12}_{11} \mid p^{12}_{00} \geq 0, p^{12}_{01} \geq 0,p^{12}_{10} \geq 0,p^{12}_{11} \geq 0, p^{12}_{00} + p^{12}_{01} + p^{12}_{10} + p^{12}_{11} = 1 \}.
\end{equation}
This imposes the standard conditions that probabilities are positive and sum to one. If we have the subset of $L$ given by $O = \{l_1\}$ then a probability distribution over its states is given by
\begin{equation}\label{eq:examplemargdist}
\{p^{1}_{0}, p^{1}_{1} \mid p^{1}_{0} \geq 0, p^{1}_{1} \geq 0, p^{1}_{0}+p^{1}_{1} = 1 \}.
\end{equation}
In order to be consistent the distribution expressed in \autoref{eq:examplejointdist} should be related to that of \autoref{eq:examplemargdist} via a marginalization matrix
\begin{equation}
\begin{pmatrix}
p^{1}_{0}\\
p^{1}_{1}
\end{pmatrix} = \begin{pmatrix}
1 & 1 & 0 & 0\\
0 & 0 & 1 & 1
\end{pmatrix}
\begin{pmatrix}
p^{12}_{00}\\
p^{12}_{01}\\
p^{12}_{10}\\
p^{12}_{11}
\end{pmatrix}.
\end{equation}

\section{Compatibility of distributions on \gnpm{}}\label{sec:compatibilityofgpms}
Here we provide an example of compatibility conditions on \gnpm{}. A more general mathematical characterization of these constraints is provided in \refsupp{} \autoref{secsupp:compatibilityofgpms}. When one has a non-trivial network architecture (corresponding to the \SH{} hypergraph like those in \autoref{fig:netsubnetcontext}), there will typically be more
than one way of obtaining a probability distribution on a set by
marginalizing a distribution on a larger set.  For instance, if we have
a network with three binary variables and two edges, $\{l_1,l_2\}$ and
$\{l_1,l_3\}$, then we can obtain a probability distribution on the
set $\{l_1\}$ either by marginalizing probabilities defined over
$\{l_1,l_2\}$ as was done above or by marginalizing probabilities
defined over $\{l_1,l_3\}$ to obtain
\begin{equation}
 \begin{pmatrix}
  p^{1}_{0}\\
  p^{1}_{1}
 \end{pmatrix} =
 \begin{pmatrix}
  1 & 1 & 0 & 0\\
  0 & 0 & 1 & 1
 \end{pmatrix}
 \begin{pmatrix}
  p^{13}_{00}\\
  p^{13}_{01}\\
  p^{13}_{10}\\
  p^{13}_{11}
 \end{pmatrix}.
\end{equation}
For an arbitrary choice of the quantities $p^{12}_{00}, \ldots,
p^{12}_{11}, p^{13}_{00}, \ldots, p^{13}_{11}$, there is no reason
that these two procedures should yield the same answers for $p^1_0$
and $p^1_1$.  If one requires that they do yield the same answer, then
one must impose consistency conditions.   In our example, these
conditions are as follows:
\begin{eqnarray}
 p^{12}_{00} + p^{12}_{01} &=&
  p^{13}_{00} + p^{13}_{01}\\
 p^{12}_{10} + p^{12}_{11} &=&
  p^{13}_{10} + p^{13}_{11}
\end{eqnarray}

More generally, given a hypergraph $\mathcal{G}$, we will be
interested in two types of consistency conditions.  We will say that a
collection of probabilities associated to a hypergraph is
\emph{locally consistent} if, whenever two hyperedges share a subset in
common, the probabilities for that subset obtained by marginalizing
the probabilities associated to one of the hyperedges will agree with
those obtained by marginalizing the probabilities associated to the
other hyperedge.  In our example above, there were only two hyperedges
present, so the conditions we exhibited constitute the entirety of the
local consistency conditions for that hypergraph.  We will denote the
set of all locally consistent probability distribution associated to a
hypergraph $\mathcal{G}$ as $\mathbb{L}(\mathcal{G})$.

We will say that a collection of probabilities associated to a
hypergraph is \emph{globally consistent} if there exists a joint
probability distribution on the totality of variables associated to
the hypergraph such that the probabilities associated to any hyperedge
are marginals of that joint distribution.  In terms of our example,
that would mean that there exist probabilities $p^{123}_{000},
p^{123}_{001}, \ldots, p^{123}_{111}$ such that the following
conditions hold:
\begin{equation}
 \begin{pmatrix}
  p^{12}_{00}\\
  p^{12}_{01}\\
  p^{12}_{10}\\
  p^{12}_{11}\\
  p^{13}_{00}\\
  p^{13}_{01}\\
  p^{13}_{10}\\
  p^{13}_{11}
 \end{pmatrix} =
 \begin{pmatrix}
  1 & 1 & 0 & 0 & 0 & 0 & 0 & 0\\
  0 & 0 & 1 & 1 & 0 & 0 & 0 & 0\\
  0 & 0 & 0 & 0 & 1 & 1 & 0 & 0\\
  0 & 0 & 0 & 0 & 0 & 0 & 1 & 1\\
  1 & 0 & 1 & 0 & 0 & 0 & 0 & 0\\
  0 & 1 & 0 & 1 & 0 & 0 & 0 & 0\\
  0 & 0 & 0 & 0 & 1 & 0 & 1 & 0\\
  0 & 0 & 0 & 0 & 0 & 1 & 0 & 1
 \end{pmatrix}
 \begin{pmatrix}
  p^{123}_{000}\\
  p^{123}_{001}\\
  p^{123}_{010}\\
  p^{123}_{011}\\
  p^{123}_{100}\\
  p^{123}_{101}\\
  p^{123}_{110}\\
  p^{123}_{111}\\
 \end{pmatrix}.
\end{equation}
We will denote the set of all globally consistent probability
distribution associated to a hypergraph $\mathcal{G}$ as
$\mathbb{M}(\mathcal{G})$.

Because marginalizing from a set of random variables to a smaller set
of variables can be accomplished by first marginalizing to an
intermediate set and then marginalizing from the intermediate set down
to the smaller set, it follows that global consistency implies local
consistency.  We will now see what conditions are needed in
addition to local consistency to ensure global consistency.

As in our example, we can express marginalization from the set $L$ of
all variables down to a hypergraph $\mathcal{G}$ in the form $v = Gx$
where $x$ is a vector whose components are probabilities associated to
$L$, $v$ is a vector whose components are probabilities associated to
$\mathcal {G}$, and $G$ is a suitable matrix.  The consistency
conditions can be expressed in terms of the fundamental spaces (kernel
and cokernel) associated to this matrix \cite{Strang1993}.  In order for a vector $v$ to
be expressible as $Gx$ for some $x$, we must satisfy the condition
that $v\cdot u = 0$ for all $u \in \mathrm{coker} (G)$.  In our
example, the cokernel of the matrix is spanned by the following two
row vectors:
\begin{eqnarray}
 &\begin{pmatrix}
  1 & 1 & 0 & 0 & -1 & -1 & 0 & 0
 \end{pmatrix}\\
 &\begin{pmatrix}
 0 & 0 & 1 & 1 & 0 & 0 & -1 & -1
 \end{pmatrix}
\end{eqnarray}
This leads to the conditions
\begin{eqnarray}
 p^{12}_{00} + p^{12}_{01} - p^{13}_{00} - p^{13}_{01} &=& 0 \\
 p^{12}_{10} + p^{12}_{11} - p^{13}_{10} - p^{13}_{11} &=& 0 .
\end{eqnarray}
Note that these are precisely the local consistency conditions which we
exhibited earlier.  It can be shown that the condition that $u \cdot v
= 0$ for all $u \in \mathrm{coker} (G)$ will always be exactly the
local consistency conditions, \refsupp{} \ref{secsupp:compatibilityofgpms}.

To obtain the global consistency conditions, we note that, if $v =
Gx$, then we also have $v = Gy$ for any vector $y$ such that $x-y$
lies in the kernel of $G$.  Choose a subspace $T$ of column vectors
which is transverse to $\mathrm{ker}(G)$ such that the union of $T$
and $\mathrm{ker}(G)$ span the column space.  Then the equation $v =
Gx$ has a unique solution if we restrict $x$ to lie in $T$.  In order
for a column vector to represent a legitimate probability
distribution, its components must all be non-negative.  Hence, we
conclude that $v$ being globally consistent is equivalent to the
following system of equations and inequalities having a solution:
\begin{equation}
\begin{aligned}\label{eq:globalconsistencyconditions}
 v &= Gx \\
 x &\in T \\
 x - y &\in \mathrm{ker}(G) \\
 y &\ge 0
\end{aligned}
\end{equation}

By using a method, such as Fourier-Motzkin elimination, to remove redundant inequalities, one can eliminate the quantities $x$ and $y$ from this
system to obtain inequallities involving only the components of $v$.
These are the global consistency conditions.

In our example, $\mathrm{ker}(G)$ is spanned by the folllowing two
column vectors:
\begin{equation}
 \begin{pmatrix}
 1 \\ -1 \\ -1 \\ 1 \\ 0 \\ 0 \\ 0 \\ 0
 \end{pmatrix}
 \begin{pmatrix}
 0 \\ 0 \\ 0 \\ 0 \\ 1 \\ -1 \\ -1 \\ 1
 \end{pmatrix}
\end{equation}
As our transverse space $T$, we will choose the space spanned by the
following basis:
\begin{equation}
 \begin{pmatrix}
  1 \\ 0 \\ 0 \\ 0 \\ 0 \\ 0 \\ 0 \\ 0
 \end{pmatrix}
 \begin{pmatrix}
  0 \\ 1 \\ 0 \\ 0 \\ 0 \\ 0 \\ 0 \\ 0
 \end{pmatrix}
 \begin{pmatrix}
  0 \\ 0 \\ 1 \\ 0 \\ 0 \\ 0 \\ 0 \\ 0
 \end{pmatrix}
 \begin{pmatrix}
  0 \\ 0 \\ 0 \\ 0 \\ 1 \\ 0 \\ 0 \\ 0
 \end{pmatrix}
 \begin{pmatrix}
  0 \\ 0 \\ 0 \\ 0 \\ 0 \\ 1 \\ 0 \\ 0
 \end{pmatrix}
 \begin{pmatrix}
  0 \\ 0 \\ 0 \\ 0 \\ 0 \\ 0 \\ 1 \\ 0
 \end{pmatrix}
\end{equation}
With this choice, the condition $x \in T$ reduces to the equations
$x_4 = x_8 = 0$.  The conditions $x - y \in \mathrm{ker}(G)$ then
become
\begin{eqnarray}
 y_1 - x_1 = x_2 - y_2 = x_3 - y_3 &=& y_4 \\
 y_5 - x_5 = x_6 - y_6 = x_7 - y_7 &=& y_8
\end{eqnarray}
If we solve these for the $x$'s, substitute the result into the
equation $v = Gx$ and eliminate the y's between the resulting
equations and the inequalities $y \ge 0$, we find the conditions $v
\ge 0$.  This, of course, is just the condition that the probabilities
be positive.  Thus, for the case of this simple hypergraph, local
consistency suffices to ensure global consistency.  In \autoref{sec:inconsistency},
we will see that this is not always the case and that the
inequalities obtained by elimination impose more conditions on the
probabilities than just positivity.

\section{Example of unsatisfiable constraints}\label{sec:inconsistency}
We will now exemplify equations and inequalities that need to be satisfied in order to guarantee the consistency conditions for the case of three variables that form the simplest nontrivial cycle where inconsistency may arise. Suppose that $L = \{l_1,l_2,l_3\}$, $P = \{0,1\}$, $\mathcal{G} = \{\{l_1,l_2\},\{l_2,l_3\},\{l_3,l_1\}\}$.

Local consistency means that the probability for the variable $l_1$ to be associated to a given state is equivalent in case we marginalize over all the other variables contained in the biological network modules of which $l_1$ is a component.  Mathematically, this reduces to two equations corresponding to the cases when the state of $l_1$ is $0$ or $1$. If we do likewise with $l_2$ and $l_3$ in place of $l_1$ we obtain the set of local consistency conditions:
\begin{equation}
\begin{aligned}\label{eq:localconsistencythreegenes}
 p^{12}_{00} + p^{12}_{01} &= p^{1}_0 = p^{13}_{00} + p^{13}_{01}, &
 p^{12}_{00} + p^{12}_{10} &= p^{2}_0 = p^{23}_{00} + p^{23}_{01}, &
 p^{13}_{00} + p^{13}_{10} &= p^{3}_0 = p^{23}_{00} + p^{23}_{10},\\
 p^{12}_{10} + p^{12}_{11} &= p^{1}_1 = p^{13}_{10} + p^{13}_{11}, &
 p^{12}_{01} + p^{12}_{11} &= p^{2}_1 = p^{23}_{10} + p^{23}_{11}, &
 p^{13}_{01} + p^{13}_{11} &= p^{3}_1 = p^{23}_{01} + p^{23}_{11}.
 \end{aligned}
 \end{equation}
These result from applying the method outlined in \autoref{sec:compatibilityofgpms} to enumerate all local consistency conditions.
Using the local consistency conditions for our example we can derive a set of inequalities that determine $\mathbb{L}(\mathcal{G})$
\begin{equation}
\begin{aligned}\label{eq:threecycinequalities}
p^{12}_{00} &= 1 + p^{12}_{11} - p^{23}_{10} - p^{23}_{11} - p^{13}_{10} - p^{13}_{11} \geq 0, \\
p^{12}_{01} &= -p^{12}_{11} + p^{23}_{10} + p^{23}_{11} \geq 0,\\
p^{12}_{10} &= -p^{12}_{11} + p^{13}_{10} + p^{13}_{11} \geq 0,\\
p^{23}_{00} &= 1-p^{23}_{10} - p^{13}_{01} - p^{13}_{11} \geq 0,\\
p^{23}_{01} &= -p^{23}_{11} + p^{13}_{01} + p^{13}_{11} \geq 0,\\
p^{31}_{00} &= 1-p^{13}_{10} - p^{13}_{01} - p^{13}_{11} \geq 0,
\end{aligned}
\end{equation}
combined with the trivial inequalities that force all probabilities to be nonnegative. Substituting the numbers from \autoref{fig:inconsistentthreecycle}A (which are $p^{12}_{00} = 0.1,\, p^{12}_{01} = 0.4,\, p^{12}_{10} = 0.4,\,p^{12}_{11} = 0.1,\,p^{13}_{00} = 0.4,\, p^{13}_{01} = 0.1,\, p^{13}_{10} = 0.1,\,p^{13}_{11} = 0.4,\,p^{23}_{00} = 0.4,\, p^{23}_{01} = 0.1,\, p^{23}_{10} = 0.1,\,p^{23}_{11} = 0.4 $) into \autoref{eq:threecycinequalities}, demonstrates that the local conditions are satisfied.

The global consistency conditions form an underdetermined system of linear equations for the putative global distribution so their solution will assume the form of a linear subspace.  The following equations arise as a result of eliminating $x$ from the equations determined by the conditions $v=Gx$, $x \in T$, $x-y \in ker \mathbf{G}$:
\begin{equation}
\begin{aligned}\label{eq:globalpositivityeqs}
p^{123}_{001} &= p^{12}_{00} - p^{123}_{000} \\
p^{123}_{010} &= p^{13}_{00} - p^{123}_{000} \\
p^{123}_{100} &= p^{23}_{00} - p^{123}_{000} \\
p^{123}_{110} &= p^{23}_{10} - p^{13}_{00} + p^{123}_{000} \\
p^{123}_{011} &= p^{13}_{01} - p^{12}_{00} + p^{123}_{000} \\
p^{123}_{101} &= p^{12}_{10} - p^{23}_{00} + p^{123}_{000} \\
p^{123}_{111} &= 1 - p^{12}_{00} - p^{13}_{00} - p^{23}_{00} - p^{123}_{000}
\end{aligned}
\end{equation}
The remaining condition $y \geq 0$ from \autoref{eq:globalconsistencyconditions} states that all the probabilities $p^{123}_{ijk}$ must be positive numbers, which is only possible if the putative marginals satisfy suitable inequalities given by
\begin{equation}
\begin{aligned}\label{eq:globalpositivityineqs}
p^{123}_{000} &\geq min(p^{12}_{00},\, p^{13}_{00},p^{23}_{00},\, 1 - p^{12}_{00} - p^{13}_{00} - p^{23}_{00}),\\
 p^{123}_{000} &\geq max(0,\, p^{13}_{00}-p^{23}_{10},\, p^{12}_{00}-p^{13}_{01},\, p^{23}_{00}-p^{12}_{10}).
\end{aligned}
\end{equation}
A minimal set of inequalities is then expressed by substituting the equalities from \autoref{eq:threecycinequalities} into the inequalities determined by \autoref{eq:globalpositivityineqs} and eliminating redundancies resulting in
\begin{equation}
\begin{aligned}\label{eq:threecycbooleinequalities}
p^{12}_{11} - p^{23}_{11} + p^{13}_{01} \geq 0, \\
1 + p^{12}_{11} - p^{23}_{10} - p^{13}_{10} - p^{13}_{01} - p^{13}_{11} \geq 0, \\
-p^{12}_{11} + p^{23}_{10} + p^{13}_{11} \geq 0, \\
-p^{12}_{11} + p^{23}_{11} + p^{13}_{10} \geq 0.
\end{aligned}
\end{equation}
The inequalities from \autoref{eq:threecycinequalities} and \autoref{eq:threecycbooleinequalities} combined with the nonnegativity inequalities together determine the global polytope $\mathbb{M}(\mathcal{G})$. For the example given in \autoref{fig:inconsistentthreecycle}A, the first of the inequalities in \autoref{eq:threecycbooleinequalities} is demonstrated to be unsatisfied in \autoref{eq:threecycboolenumbers}
\begin{equation}
\begin{aligned}\label{eq:threecycboolenumbers}
0.1 - 0.4 + 0.1 \not\geq 0, \\
1 + 0.1 - 0.1 - 0.1 - 0.1 - 0.4 \geq 0, \\
-0.1 + 0.1 + 0.4 \geq 0, \\
-0.1 + 0.4 + 0.1 \geq 0.
\end{aligned}
\end{equation}
This indicates that data consistent with \autoref{fig:inconsistentthreecycle}A could not derive from the network depicted there.

\section{Cyclic network contexts can impose unsatisfiable constraints}\label{sec:cycliccontextunsatisfiableconstraints}

Each node of the \SH{} graph in \autoref{fig:inconsistentthreecycle}A can be associated to the probability distribution that specifies probabilities for each biological variable to be observed in each of the states determined by the coarse-graining process described in \autoref{sec:genenetworkphenmap}. Each edge of the graph specifies a joint probability distribution for both of the nodes it contains (or connects) to simultaneously take on a given pair of values. Note that this does not imply the existence or absence of a physical interaction between the variables represented by these two nodes. Together, these probabilities represent constraints that the network context may impose upon the network.
We assume three variables are observed via all possible pairwise combinations and that via the coarse-graining process we have binned the state of each variable into one of two classes.
Each node of the graph in \autoref{fig:inconsistentthreecycle}A represents a probability distribution over the observation of each variable in either of the two states established in the coarse-graining process.
Each of the probability tables adjacent to each edge in the graph assigns a probability distribution to the set of maps from the nodes connected by the edge to all possible combinations of the network states. As these maps take collections of biological network variables as input and produce collections of network states as outputs we refer to them as \gnpm{} and thus to the associated probability distributions as probability distributions over \gnpm{}.

Suppose the normalized contingency tables in \autoref{fig:inconsistentthreecycle}A are meant to represent the ostensible structure and parameters of a biological process. It is often necessary to attempt to infer the parameters of such a model from data under the assumption that the structure of a given network architecture falls within the model class defined by a given graph. \autoref{fig:inconsistentthreecycle}B represents a case in which a hypothetical dataset is consistent with its derivation from a joint probability distribution whereas \autoref{fig:inconsistentthreecycle}C represents a case of inconsistency where the pairwise distributions are each individually consistent distributions, but, together, the three pairwise distributions are not consistent with any joint distribution over the states of all three network variables.
This inconsistency is made possible by the fact that the network architecture in \autoref{fig:inconsistentthreecycle}A contains a cycle \cite{Lauritzen1996,Geiger2006,Wainwright2007} and that we have given an ostensible data set leading to the inference of parameters that could not possibly derive from a joint probability distribution over all three network variables.

If this situation arises, it indicates some systematic error in the transfer of information whether it occurs intrinsically to the system wherein a network has inconsistent constraints placed upon it by its network context or as part of the scientific data collection process. In the former case, this can be resolved by modifying the inconsistent constraints in such a manner that they become consistent with or without modifying the network architecture in doing so. In the latter case, this may result from employing a model which 1) takes insufficient account of the network context and 2) relies on coarse-grained observations. In either case, the synthetic gene circuit schematized in \autoref{fig:condgenescenario} serves as one mechanism implementing the example presented in \refsupp{} \autoref{secsupp:apparentinconsistency}. It consists of four genes each of which is capable of taking on three different states \cite{Rieckh2013a}. However, observing two out of the three states measured pairwise from three out of the four genes could result in data that would appear to be inconsistent. Such an observation would demonstrate without having to have knowledge of the correct network architecture, that the current model is insufficient to represent the underlying process.

\begin{figure}[!ht]
\centering
\noindent\includegraphics[width=0.9\columnwidth]{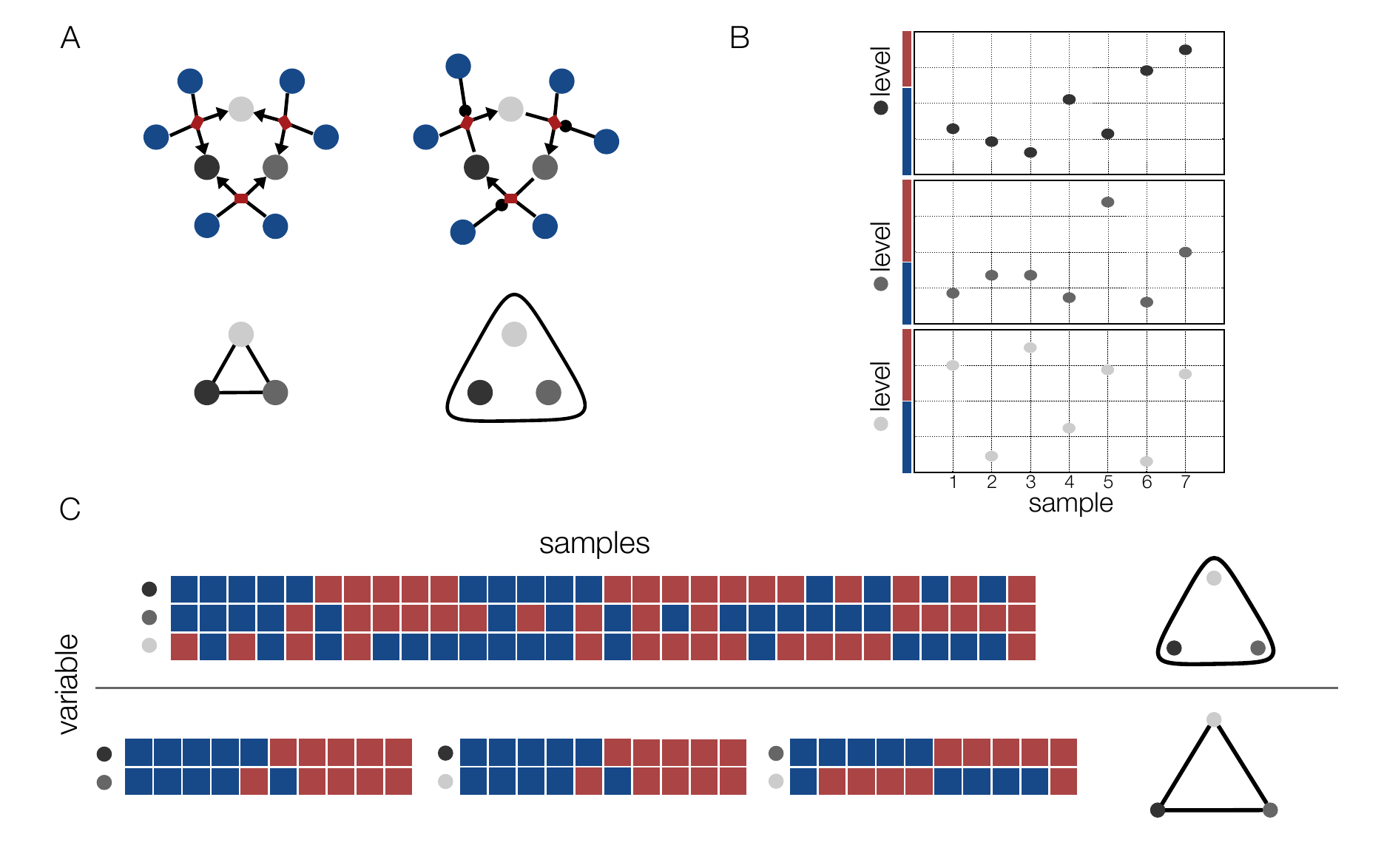}
\caption{{\bf Coarse-graining of biological network data.} (A) SBGN (top) and \SH{} (bottom) representation of two different biological networks. (B) Example binary coarse-graining of biological network data. For each sample a measurement is taken for all three variables in the focal subnetwork. The levels are binned into one of two classes represented by the red---- and blue bars representing relatively high and low levels respectively. (C) Heat map representation of coarse-grained data under the assumption of two different network architectures. The samples on top and the associated measurement structure correspond to the case where constraints are placed on all three variables by a single element of the network context (\autoref{fig:stochdynscheme} top row). The bottom represents the case where all three pairs are each independently constrained by elements of the network context (\autoref{fig:stochdynscheme} bottom row).}
\label{fig:expression_concept}
\end{figure}

For the case of the architecture in \autoref{fig:inconsistentthreecycle}A, and moreover for any network architecture of any size that contains one or more cycles, the possibility of finding a joint distribution over all network variables that satisfies all constraints capable of being imposed upon it requires the implicit assumption that the structure of the network context can be viewed simultaneously as that of \autoref{fig:expression_concept}C top and that of \autoref{fig:expression_concept}C bottom. The spaces of probability distributions corresponding to the constraints that can be imposed upon the two network acrhitectures contrasted in \autoref{fig:expression_concept}C are different. We can now apply the process described in \autoref{sec:compatibilityofgpms} to classify the geometries and thus relationships among the spaces of probability distributions associated to constraints that can be imposed on all possible network architectures with a given number of variables.

\section{Geometry of probabilistic constraints on network states}\label{sec:probconstrgeometry}
\begin{figure}[!ht]
\centering
\noindent\includegraphics[width=0.9\columnwidth]{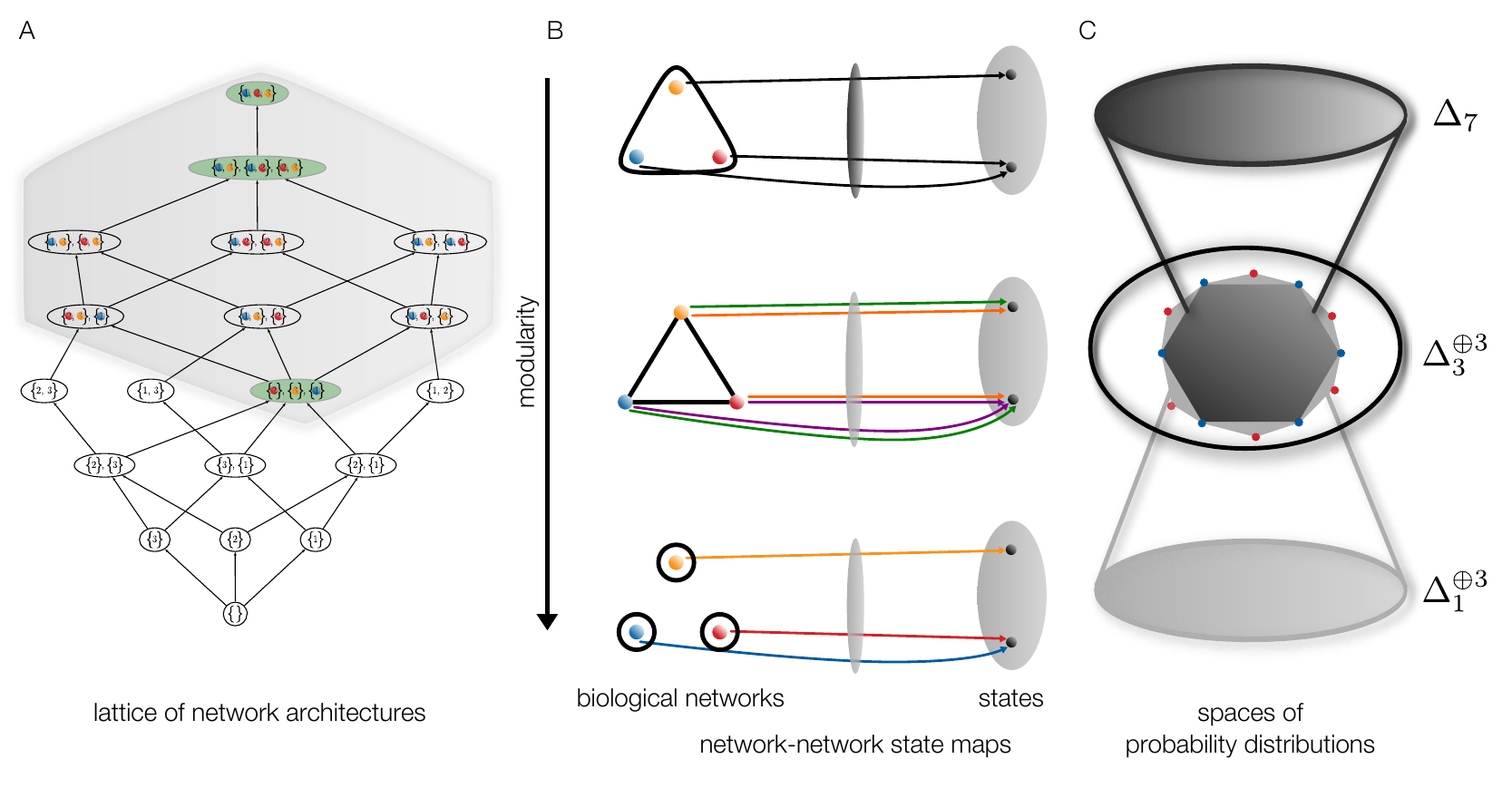}
\caption{{\bf Relationship between biological network models and spaces of probability distributions.} (A) The collection of all possible network architectures over three variables forms a lattice represented here by its Hasse diagram. An analogous lattice of network architectures exists for any number of variables. The Hasse diagram shows the manner in which network architectures are hierarchically related and are thus able to be embedded within one another. (B) Explicit examples of \gnpm{} over three network architectures from panel A highlighted in green are represented as arrows mapping the variables represented as nodes of the graph underlying the network architecture into the collection of network state values determined by the coarse-graining chosen in \autoref{fig:expression_concept}B. There is a different collection of possible \gnpm{} depending upon the structure of the network architecture. (C) Each collection of \gnpm{}, one representative for each network architecture depicted in panel B, is associated to a space of probability distributions defined over it. Moreover, the spaces of probability distributions associated to each graph are related via marginalization maps. The top level represents a joint probability distribution (i.e. $\Delta_7$: the eight-dimensional probability simplex) which can be marginalized to the middle space (i.e. $\Delta_3^{\oplus 3}$: the union of three copies of the four-dimensional probability simplex) which in turn can be marginalized to the bottom space (i.e. $\Delta_1^{\oplus 3}$: the union of three copies of the two-dimensional probability simplex). The light gray polytope in the middle, $\mathbb{L}(\mathcal{G})$, represents the space of distributions consistent with the marginalization map from the middle to the bottom. The dark gray polytope, $\mathbb{M}(\mathcal{G})$, represents the space of probability distributions consistent with marginalization from the top to the middle.}
\label{fig:conediagram}
\end{figure}

The relationships among possible network architectures are given by the lattice, which in this case indicates ordering by subset inclusion, of reduced subsets of biological network variables (i.e. collections of subsets of variables where no subset in the collection is a subset of another one, \autoref{sec:networkcontext} and \refsupp{} \autoref{sec:covergenotypespace}). For example, \autoref{fig:conediagram}A shows the lattice of reduced subsets of three variables. We are only interested in those subsets that contain at least one instance of each variable. Restricting to the subsets of variables satisfying this condition corresponds to the region highlighted with a gray background in \autoref{fig:conediagram}A. Each network architecture corresponds to a different modularization of the \gnpm{} by the network context. For example, \autoref{fig:conediagram}B shows in the same vertical order the different maps induced by the three architectures highlighted in green in \autoref{fig:conediagram}A.

We consider those network architectures found lower in the lattice of \autoref{fig:conediagram}A to be of higher modularity because each corresponds to the increasing restriction from placing constraints on higher- to placing constraints on lower-order correlations among variables. \autoref{fig:conediagram}B top corresponds to the least modular network architecture because constraints are placed upon correlations among all three variables. \autoref{fig:conediagram}B middle exhibits an elevated degree of modularity because constraints are placed upon correlations among pairs of variables. Similarly, \autoref{fig:conediagram}B bottom is even more modular because constraints are placed upon each variable individually.

Each of the network architectures in \autoref{fig:conediagram}A can be associated to a pair of spaces of probability distributions over \gnpm{}. These correspond to the spaces of globally, $\mathbb{M}(\mathcal{G})$, and locally, $\mathbb{L}(\mathcal{G})$, consistent distributions described in \autoref{sec:compatibilityofgpms} and \autoref{secsupp:compatibilityofgpms}. \autoref{fig:conediagram}C schematically depicts the relationships among the probability distributions associated to the corresponding architectures and \gnpm{} in \autoref{fig:conediagram}B. For \autoref{fig:conediagram}C top, $\mathbb{M}(\mathcal{G}) = \mathbb{L}(\mathcal{G})$. The inconsistency noted in the previous section between the architectures \autoref{fig:conediagram}B top and middle is a result of the differing geometries in \autoref{fig:conediagram}C middle. There, the smaller darker gray region, $\mathbb{M}(\mathcal{G})$, defined by the inequalities expressed in \autoref{eq:threecycinequalities} and \autoref{eq:threecycbooleinequalities} corresponds to the space of probability distributions defined over all possible \gnpm{} associated to the network architecture in \autoref{fig:conediagram}B middle. Similarly, the lighter gray region defined by \autoref{eq:threecycinequalities} alone corresponds to $\mathbb{L}(\mathcal{G})$ for \autoref{fig:conediagram}B middle and thus $\mathbb{M}(\mathcal{G}) < \mathbb{L}(\mathcal{G})$ in the latter case.

\section{Naive likelihood of sampling unsatisfiable constraints}\label{sec:volrat}
\begin{figure}[!ht]
\centering
\noindent\includegraphics[width=1.0\columnwidth]{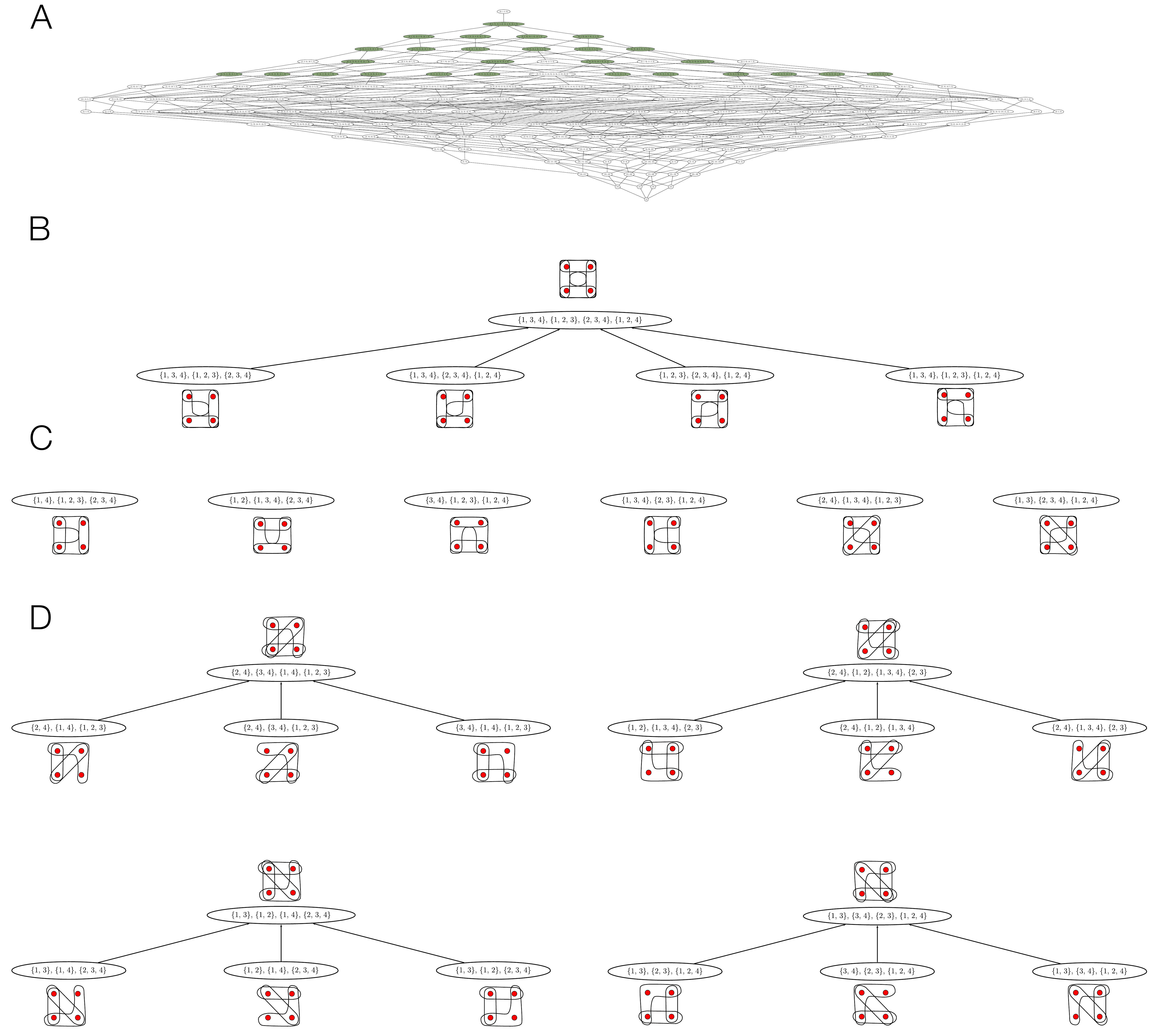}
\caption{{\bf Hierarchical relationships among all possible classes of hypergraphs that are not graphs (i.e. not 2-uniform) but have cycles.} (A) There is a Hasse diagram for the lattice of network architectures analogous to that of \autoref{fig:conediagram}A but defined on four rather than only three variables. Within this lattice some of the graphs have cycles and some do not. (B) The highest levels of the Hasse diagram associated to the lattice of network architectures on four variables containing hypergraphs having cycles. (C) and (D) contain lower levels of network architectures containing cycles. Each of the four panels in (D) are on the same level. In total, each level represents an isomorphism class of hypergraphs. Therefore, there are five isomorphism classes of non-2-uniform hypergraphs representing network architectures on four variables that contain cycles leading to the relationship between spaces of probability distsributions on associated genotype-phentoype maps analogous to that of \autoref{fig:conediagram}C.}
\label{fig:non2uniformcyclichypergraphhasse}
\end{figure}

Relationships between spaces of potential constraints placed upon patterns of network states like that of \autoref{fig:conediagram}C middle occur for all network architectures defined over any number of variables so long as there exists at least one cycle in the corresponding network architecture, \autoref{sec:cycliccontextunsatisfiableconstraints}. For the case of three variables, there is only one class of graphs containing a cycle, which is that of \autoref{fig:conediagram}B middle. For the case of four variables there are nine different classes of hypergraphs containing cycles and these nine classes can be split into two groups depending upon whether or not the edges of the graphs are each restricted to represent correlations among only two variables. \autoref{fig:non2uniformcyclichypergraphhasse} shows the components of the analogous lattice to that of \autoref{fig:conediagram}A as well as these different classes of network architectures on four variables having cycles.

Given this larger collection of network architectures with cycles we can assess the relative sizes of the spaces $\mathbb{M}(\mathcal{G})$ and $\mathbb{L}(\mathcal{G})$ (\autoref{fig:conediagram}C middle) of probability distributions over \gnpm{}. We assess the likelihood of choosing a point in $\mathbb{M}(\mathcal{G})$ at random by computing the ratio of the volume of $\mathbb{M}(\mathcal{G})$ (associated to the non-modular network architectures analogous to that of \autoref{fig:conediagram}B top with a single edge containing all four variables), whose architecture and thus volume is fixed, to that of $\mathbb{L}(\mathcal{G})$, whose volume varies according to each of the cyclic graphs associated to a network architecture on four variables. We refer to this number as the global:local volume ratio or~$\frac{\text{Vol}(\mathbb{M}(\mathcal{G}))}{\text{Vol}(\mathbb{L}(\mathcal{G}))}$~(see \autoref{sec:compatibilityofgpms} and \refsupp{} \autoref{secsupp:compatibilityofgpms} and \autoref{secsupp:fourcycleexample}). The comparison defined by this ratio is meaningful since $\mathbb{L}(\mathcal{G})$, \autoref{eq:localpolytope}, and $\mathbb{M}(\mathcal{G})$, \autoref{eq:globalpolytope} are of the same dimension. In the case where the constraints defining $\mathbb{L}(\mathcal{G})$ are eliminated, the analog of this volume ratio would be $0$ for all $\mathcal{G}$. This volume ratio determines the \emph{a priori} likelihood of observing inconsistency for a given network architecture. The consistency check involved in computing this ratio can be used as a test demonstrating, for those cases exhibiting inconsistency, that the model being used is incorrect in the sense that it does not correspond sufficiently to the actual network context determining the constraints placed upon the network. Consider the probability of locally versus globally consistent \emph{observations} ($p(\mathbb{L}(\mathcal{G})_O)$  vs $p(\mathbb{M}(\mathcal{G})_O)$ respectively) separately from the probability of locally versus globally consistent \emph{models} ($p(\mathbb{L}(\mathcal{G})_M)$  vs $p(\mathbb{M}(\mathcal{G})_M)$ respectively) that accurately reflect the underlying process. We can then estimate the probability of having a locally consistent model despite obtaining globally consistent observations, $p(\mathbb{L}(\mathcal{G})_M | \mathbb{M}(\mathcal{G})_O)$, via a simple application of Bayes' theorem
$$
p(\mathbb{L}(\mathcal{G})_M | \mathbb{M}(\mathcal{G})_O) = \frac{p(\mathbb{M}(\mathcal{G})_O | \mathbb{L}(\mathcal{G})_M)p(\mathbb{L}(\mathcal{G})_M)}{p(\mathbb{M}(\mathcal{G})_O | \mathbb{L}(\mathcal{G})_M)p(\mathbb{L}(\mathcal{G})_M) + p(\mathbb{M}(\mathcal{G})_O | \mathbb{M}(\mathcal{G})_M)p(\mathbb{M}(\mathcal{G})_M)},
$$
where $p(\mathbb{M}(\mathcal{G})_O | \mathbb{M}(\mathcal{G})_M)=1$, the volume ratio described above corresponds to $p(\mathbb{M}(\mathcal{G})_O | \mathbb{L}(\mathcal{G})_M)$, and one could consider the impact of different prior probabilities, $p(\mathbb{L}(\mathcal{G})_M)$, of having a locally consistent model.

\autoref{fig:ncycvolrat}A and B shows the results of computations of this global:local volume ratio for fourteen different hypergraphs. \autoref{fig:ncycvolrat}C and D shows the dimension of the spaces within which these volumes are computed. The spaces are equivalent and thus the volume ratio equal to one for graphs lacking cycles (e.g. the first three graphs along the $x$-axis of \autoref{fig:ncycvolrat}A). For the nine network architectures in \autoref{fig:ncycvolrat}A and B containing cycles, the volume ratio is strictly less than one. This quantifies the probability that the network architecture depicted along the $x$-axis will be able to satisfy the constraints that the associated network context is capable of placing upon it.

\section{Potential for unsatisfiable constraints may bias the sampling of network architectures by evolutionary processes}\label{sec:unsatisfiableconstrevolution}

\begin{figure}[!ht]
\centering
\noindent\includegraphics[width=0.9\columnwidth]{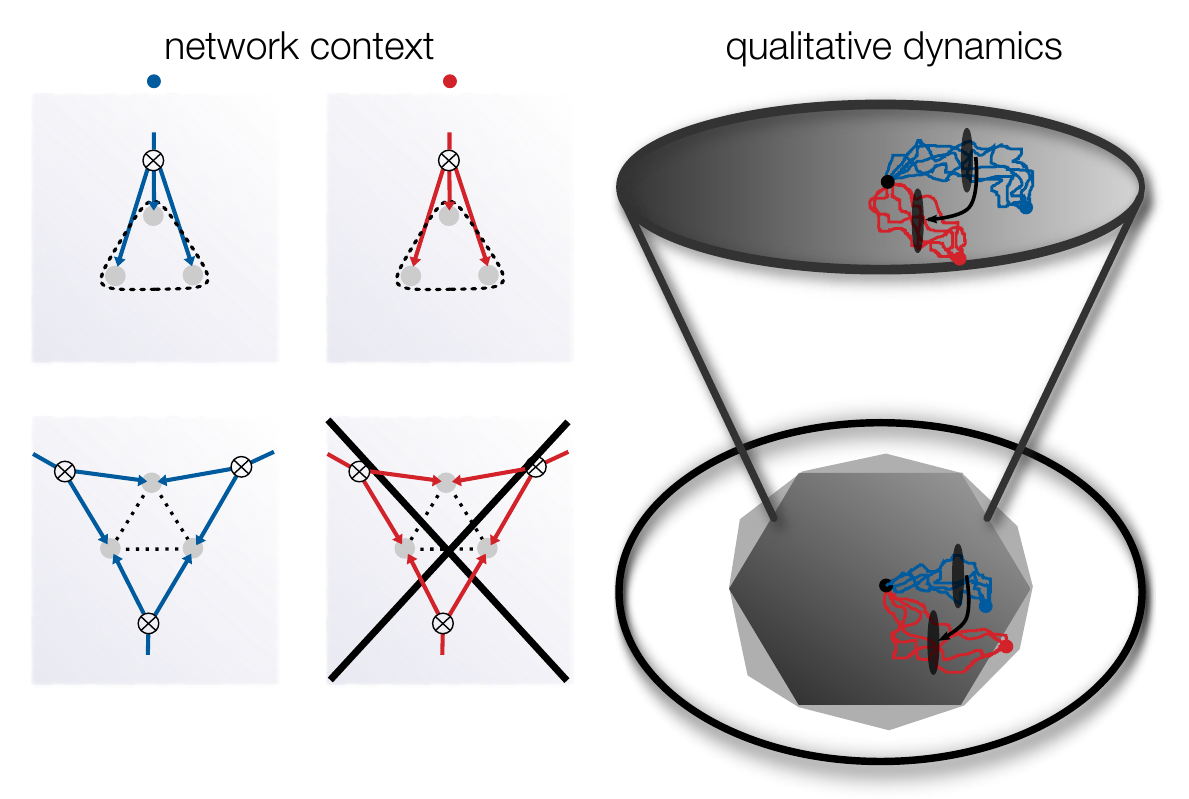}
\caption{{\bf Constraints imposed on stochastic biological networks and evolutionary dynamics by network architecture.} Schematic representation of a potential network context (left) for each of the hypothetical stationary probability distributions associated to the fitness peak established by the blue and red points within the spaces of probability distributions represented on the right.
Either of the two network architectures represented on the left are capable of achieving the stationary distribution over \gnpm{} specified by the blue stationary distribution associated to a hypothetical fitness peak. On the other hand, only the network architecture from the top (and not the bottom) is capable of achieving the red stationary distribution representing an alternative potential fitness peak.}
\label{fig:stochdynscheme}
\end{figure}

The satisfiability of constraints capable of being placed on the various architectures is logically a function of whether or not the network architecture is cyclic or acyclic. For those network architectures containing cycles, there are certain functional requirements that can be achieved so long as only local and not global consistency is required of them. Once global consistency is imposed as in the structure corresponding to the joint correlations among all variables, those functions that were accessible when only local consistency was imposed are unavailable. For acyclic network architectures, there is no difference between the satisfiability of locally or globally imposed constraints. \autoref{fig:stochdynscheme} right shows a schematic of one potential scenario by which a given cyclic network architecture may be selected against. The black points in the center represent an initial condition of a stochastic process that is selected for its ability to achieve one of two different stationary distributions represented by the blue and the red points respectively. This is equivalent to placing a fitness landscape given by a function whose maximum is located at the given points and defined over the relevant space of probability distributions. The network architecture represented in the top row of \autoref{fig:stochdynscheme} is able to achieve as its stationary distribution any of the constraints capable of being imposed upon it that are consistent with its architecture because it is acyclic. On the other hand, the network architecture in the bottom row is incapable of achieving certain constraints that may be imposed upon it by a network context consistent with its architecture because it is cyclic.

When selective pressure is induced equivalent to the distribution located at the blue point, or at any other point within the dark gray region, either of the architectures are essentially equivalent with respect to the statistics of samples from their corresponding probability distributions and they can thus be considered as members of an evolutionarily neutral space. On the other hand, selective pressure equivalent to the probability distributions located at the red point differentiates between the networks of the top and bottom row or equivalently between the network of the bottom row when global consistency is imposed versus the same network when only local consistency conditions are imposed. The same qualitative relationship holds true for the spaces of probability distributions of all network architectures of any size and for any number of different levels in the discrete coarse-graining of network states so long as the graph associated to the relevant correlations among variables contains at least one cycle.

The distinction between cyclic and acyclic network architectures with respect to the ability to have unsatisfiable constraints placed upon them is sharp. However, within the class of cyclic network architectures, the likelihood of having unsatisfiable constraints imposed on a given network architecture increases, at least approximately, with the number of cycles in the given network architecture (\autoref{fig:ncycvolrat} and \autoref{sec:volrat}). This indicates that the strength of selection against network architectures with a larger number of nested cycles is likely to be stronger than that against network architectures with a relatively smaller number of cycles. Initiating an evolutionary process with a large network containing many nested cycles may then result in the elimination of some via any process that can result in cycle breakage until the number of nested cycles decreases sufficiently so that the intrinsic strength of selection against cycles reaches equilibrium with the rate at which new cycles form. One possibility, depending upon the overall relationship between these rates, is a hierarchical-modular one where a globally hierarchical network has a number of cyclic modules, each of whose size is small relative to the overall size of the network, interspersed throughout.

\section{Discussion}

When biological networks are studied, we remove a subnetwork from a larger context~\cite{Alon2007}. Depending upon the scale of the study, the boundary between subnetwork and network context may vary. For example, in a relatively small-scale study the subnetwork may consist of a few genes and metabolites where the context is comprised of other genes, metabolites, and intracellular structures. For relatively large-scale models attempting to take into account all of the processes comprising a single-celled organism, the network context consists of the variables in that organism's environment. In even larger-scale studies of multicellular organisms, populations, or communities the same general principle applies by appropriately shifting the boundary between the subnetwork and network context.

One salient feature applying at any scale is that the structure of the network context plays a crucial role in determining whether or not unsatisfiable constraints on the stochastic dynamical patterns of network states may arise at all. We note based on previously existing results that mutually incompatible constraints are only capable of arising when the network architecture contains a cycle. Moreover, our results suggest the likelihood of mutually incompatible constraints arising relative to network architecture increases with the number of cycles in that network architecture. An evolutionary process exhibiting uniform sampling over the space of network architectures and the space of possible constraints within each network architecture, would thus be expected to exhibit a bias toward the breakage of cycles. One would not expect such a bias to eliminate the existence of cycles in biological networks. However, it is reasonable to expect on the basis of this result a kind of hierarchical modularity: where modules that may possess cycles and are small relative to the overall size of the network exist within a globally hierarchical network structure. Of course, there are other factors which may contribute to the development of such network architectures.

\begin{figure}[!ht]
\centering
\noindent\includegraphics[width=1.0\columnwidth]{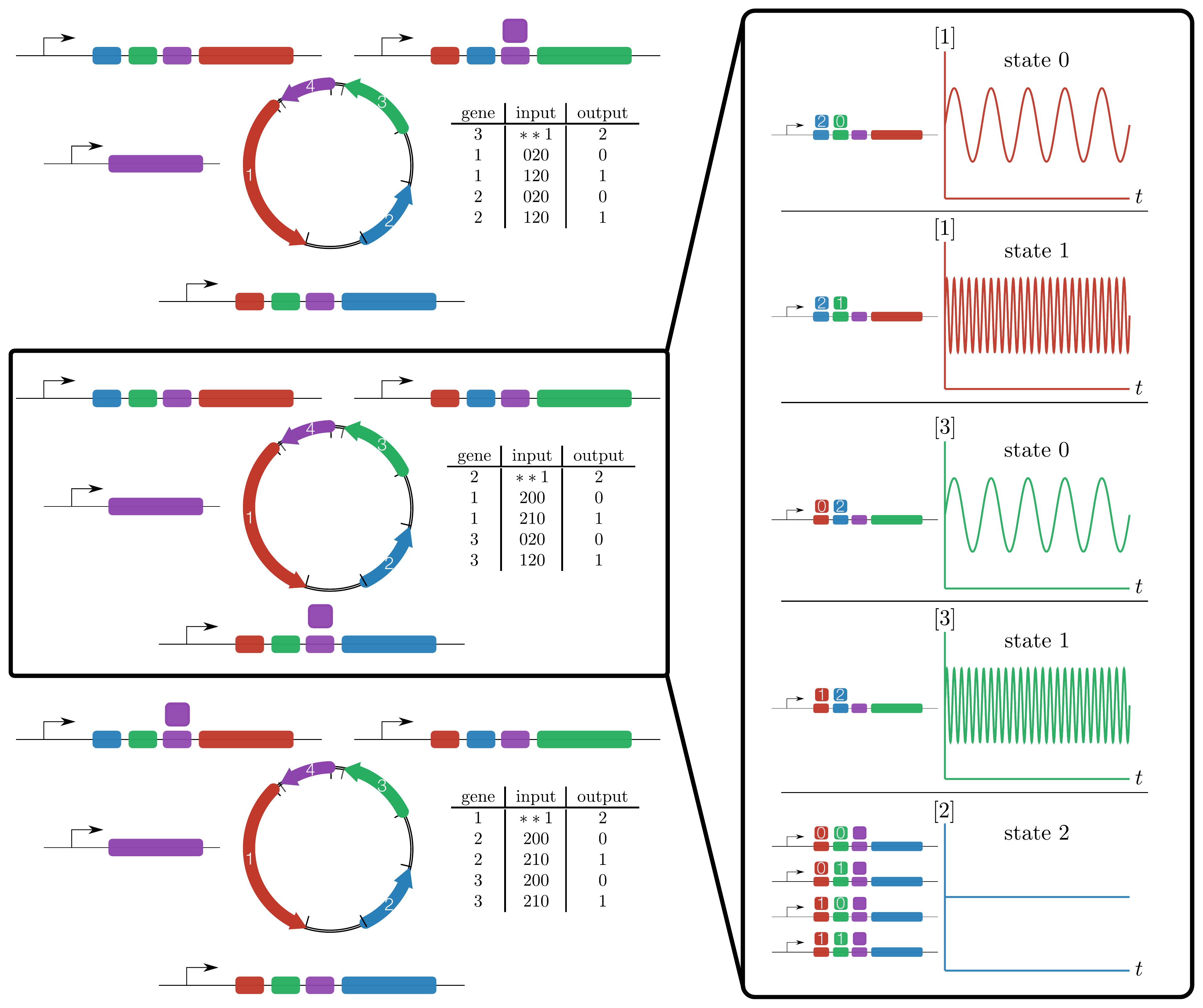}
\caption{{\bf Schematic synthetic gene circuit capable of exhibiting apparent inconsistency.} A synthetic gene circuit consisting of four genes possesses one gene (purple) whose product is assumed to be present at low copy number and binds randomly with equal affinity to operator sites existing within the operons of each of the other three genes (red, blue, and green). These latter three genes each possess operator sites for the other two, but do not possess autoregulatory operators. They also each exhibit three states represented by three dynamical modes that may involve intermediates not explicitly represented here \cite{Cai2008,Dolmetsch1998}. If the first gene is bound to the operator of another gene, the output is forced into a zero frequency infinite period, or DC, mode (state 2) regardless of the binding state of the other operators. If the first gene is unbound, then the expression state can be switched between low (state 0) and high (state 1) frequency modes depending upon the binding states of other genes as indicated. Note that operators for each of genes one to three are insensitive to the DC mode. Observing pairs of genes one to three and ignoring the state corresponding to the DC mode can lead to apparent inconsistency.}
\label{fig:condgenescenario}
\end{figure}

It will be important in future work to examine this prediction more closely in the context of developing bottom-up stochastic process models that allow for the explicit encoding and solution of models of more complex biological networks~\cite{Walczak2009,Mugler2009}. It is possible that the specific dynamics of a given network context may lead to apparent access to correlations that are otherwise inaccessible. In the case of gene-regulatory networks, this may occur via a form of cis-regulation that enables the breakage of statistical dependence in a time-dependent manner \autoref{fig:condgenescenario}. But such a scenario seems much less plausible than the ability to resolve inconsistency by breaking cycles in the network architecture. In the long term, the latter corresponds to what is observed in hierarchically organized transcription factor networks \cite{Jothi2009,Bhardwaj2010,Chalancon2012,Colm}. The mechanism outlined here is consistent with previous analyses of hierarchical modular gene regulatory network architectures~\cite{Ravasz2002,Segre2005,Wagner2007,Erwin2009,Jothi2009,Bhardwaj2010,Colm}.

To contribute to the broader goal of establishing an integrated framework that synthesizes hypothesized intrinsic and extrinsic constraints necessary to understand the functioning and evolution of biological systems, here we have traced a path from biological network architecture to network state constraint satisfiability, and, via the impact of network states on higher-level properties culminating in macroscopically observable phenotypes, to evolutionary processes. In the particular context of gene-regulatory networks, one goal of measuring gene expression at transcriptomic scale is to uncover the structure of the generative process encoded in the interactions involved, but, so far, even the most sophisticated methods of describing them at the mechanistic level are only solvable for extremely simple regulatory network architectures~\cite{Walczak2009,Mugler2009}. This fact has, in part, motivated computational biologists to develop a large collection of algorithms to infer aspects of this structure~\cite{Anastassiou2007,DeSmet2010} and experimental biologists to compare networks on the basis of their hierarchical and modular architecture~\cite{Ideker2012}. Our model and its framework put forward a class of fundamental constraints that may impact the expected structure of biological networks. The fact that the satisfiability of the space of possible constraints that can be imposed upon a network is dependent upon the structure of the network context provides a mechanism by which natural selection may exhibit a fundamental bias in its sampling of biological network architectures.

\section*{Authors' contributions}
CS developed the project. CS, XP, RSP, and AB performed research. All authors contributed to writing the paper.

\section*{Acknowledgments}
Support was provided by NIH MSTP training grant T32-GM007288 to CS and DB, the Fulbright program to XP, and NIH R01-CA164468-01 and R01-DA033788 to AB. The authors thank Jay Sulzberger for sharing important discussions and mathematical insight. We thank Noson Yanofsky and Andrew Yates for helpful discussions.

\bibliography{bib/books,bib/papers}

\pagebreak
\section*{Figure Legends}

\begin{figure}[!ht]
\centering
\noindent\includegraphics[width=0.9\columnwidth]{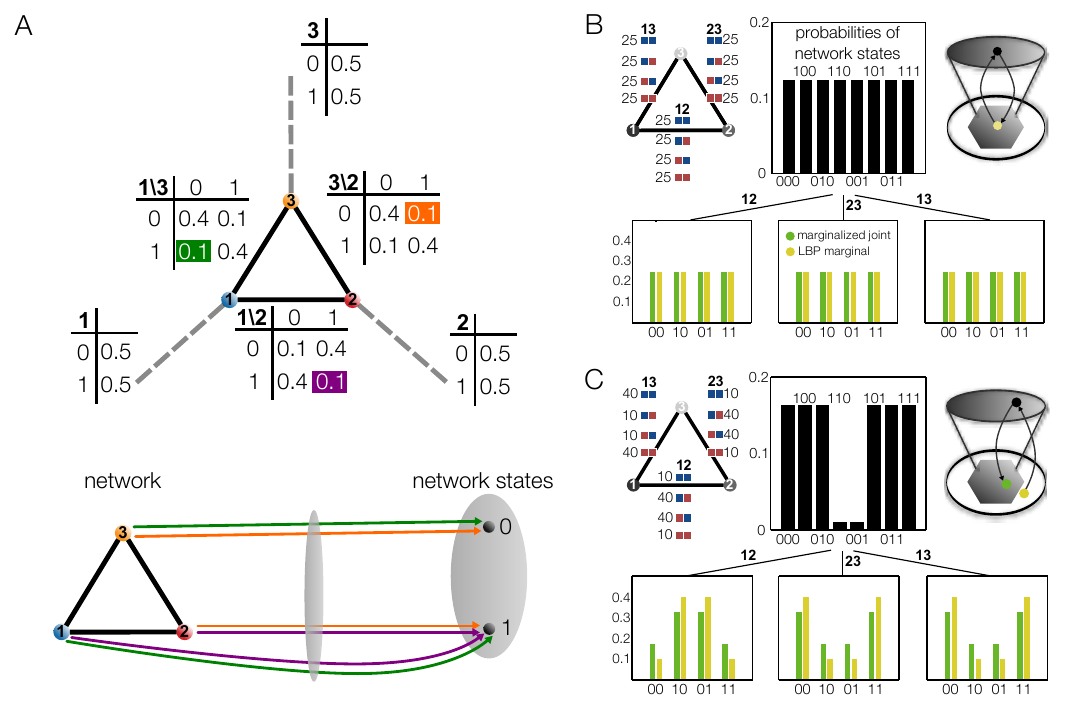}
\caption{{\bf Model of inconsistent network state data.} (A) An example structured according to the bottom row of \autoref{fig:expression_concept}C. The graph contains three nodes each representing one of the variables depicted in \autoref{fig:expression_concept}A. The dashed gray line coming from each variable points to the single variable marginal distribution depicted in the associated table. The pairwise edge marginal distributions are placed along the edges. The highlighted table entries (top) represent the constraint probabilities on the \gnpm{} represented by the equivalently colored arrows (bottom). The binary values representing variable states derive from the coarse-graining process over continuous network state data depicted in \autoref{fig:expression_concept}B. (B) (top-left) Representation of three hundred samples comprising a data set consistent with a uniform distribution over all \gnpm{} from the model in panel A. (top-middle) The joint probability distribution  given in the top-left panel. The green bars in the bottom three panels represent the marginalization of this joint distribution according to the structure of the graph. The yellow bars in the bottom three panels represent the ostensible marginal distributions determined via the sum-product algorithm (loopy belief propagation) \cite{Barber2012}. (top-right) A schematic where the top gray ellipse represents the space of joint probability distributions on three variables and the hexagon represents the pairwise marginals within their natural embedding space (see \autoref{fig:conediagram}).  For this data, maximum likelihood estimation (exact) and loopy belief propagation (approximate) yield equivalent points within the space of pairwise marginals. (C) Same as B, but with data consistent with \autoref{fig:expression_concept}C bottom, which in the limit of a large amount of data would converge to the ostensible node and edge marginal distributions in panel A. For the given data set, maximum likelihood estimation and loopy belief propagation yield different points within the natural embedding space of the pairwise marginals.}
\label{fig:inconsistentthreecycle}
\end{figure}

\begin{figure}[!ht]
\centering
\noindent\includegraphics[width=0.9\columnwidth]{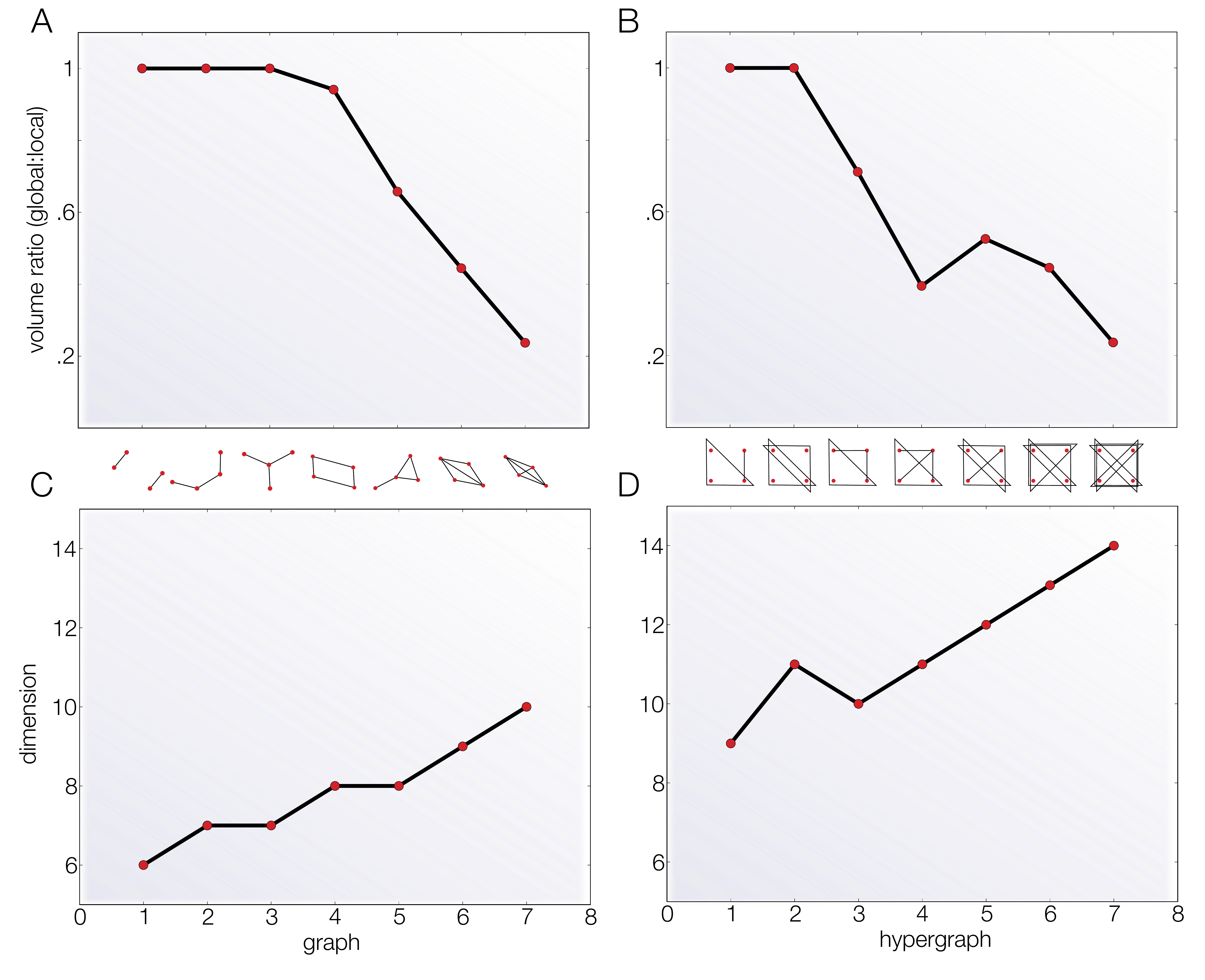}
\caption{{\bf Non-modular to modular probability space volume ratio.} (A) and (B) show the ratio $\frac{\text{Vol}(\mathbb{M}(\mathcal{G}))}{\text{Vol}(\mathbb{L}(\mathcal{G}))}$ associated to 2-regular and non-2-regular network architectures respectively. The (hyper)graph associated to each value of the volume ratio is displayed along the x-axis of each panel. (C) and (D) show the natural dimension of the space of probability distributions associated to $\mathbb{M}(\mathcal{G})$ and $\mathbb{L}(\mathcal{G})$ for each hypergraph.}
\label{fig:ncycvolrat}
\end{figure}

\FloatBarrier

\pagebreak
\beginsupplement
\setcounter{secnumdepth}{4}
{\noindent\Large\bfseries Supplementary Material}

\section{Outline}\label{secsupp:outline}
In the \refsupp{} we provide a more formal mathematical description of the results we make use of in the main text. In \autoref{sec:covergenotypespace} we characterize biological network architectures as a collection of subsets, each individually referred to as a module, of network variables that defines a hypergraph over those network variables. \autoref{secsupp:probabilitydistributionsonnetworks} provides a functorial description of probability distributions defined over such network architectures and the mappings between those network architectures and the states of the modules of the network. \autoref{secsupp:coarsegrainingphenotypes} characterizes the manner in which a hierarchy of coarse-grained network states can be viewed as a refinement of the genotype-phenotype map, where the genotype and phenotype correspond to two different levels within this hierarchy, but maps between any two levels are considered to define valid coarse-grainings. \autoref{secsupp:compatibilityofgpms} provides a sheaf theoretic formulation of the local and global consistency conditions that are logically imposed upon probability distributions over collections of such maps from some lower- to some higher-level in the hierarchy of coarse-grainings. \autoref{secsupp:fourcycleexample} complements \autoref{sec:compatibilityofgpms} of the main text providing a detailed example computation of the ratio of volumes between the polytopes corresponding to the global and local consistency conditions for the four-cycle network architecture.

\section{Biological network architecture}\label{sec:covergenotypespace}
A module of a biological network is represented by a subset of variables, $O \subseteq L$. A biological network architecture, $\mathcal{G}$, may then be represented by a subset of all possible such modules. This is to say that $\mathcal{G}$ is a subset of the set of all subsets of $L$, $\mathcal{G} \subseteq \mathcal{P}(L)$, that satisfies the following two conditions
\begin{enumerate}
\item $\cup_i O_i = \cup \mathcal{G} = L$,
\item If $O,O' \in \mathcal{G}$ and $O \subseteq O'$ then $O = O'$.
\end{enumerate}
The first condition is just a statement that $\mathcal{G}$ represents a decomposition of the collection of all variables under consideration into subsets and this is why we refer to $\mathcal{G}$ as a collection of biological network modules. The second condition means simply that we will not consider nested subsets and so we will take for our $O \in \mathcal{G}$ the biggest $O \in \mathcal{G}$ that is not a subset of some other $O' \in \mathcal{G}$. The second condition also implies that if a given subset of variables $O'$ is compatible in a sense to be explained more precisely in what proceeds then any smaller subset of variables $O$ is also compatible.

Mathematically, the two conditions given above state that $\mathcal{G}$ is a \emph{covering} of the set $L$.  This is equivalent to $(L, \mathcal{G})$ being a reduced hypergraph, Sperner family, or clutter over $L$ \cite{Lauritzen1996}.  Coverings $\mathcal{G}$ of the space of biological network variables contain the necessary information to make precise what we heuristically refer to at other points in this paper as modularity in order to cohere with standard terminology in systems biology literature while attempting to submit our own precise interpretation of the relatively colloquial concept.

\section{Functorial formulation of probability distributions over network modules}\label{secsupp:probabilitydistributionsonnetworks}

As stated in \autoref{sec:probabilitydistributionsonnetworks}, essentially all studies of biological networks consider states of subsets of variables that interact either directly or indirectly.  We will represent these modules as subsets of $L$ and their states as functions from these subsets to $P$.

The power set of $L$, which we shall denote as $\mathcal{P}(L)$, can be regarded  as a category~\cite{Lane1998,MacLane1992,Awodey2006,Abramsky2011} in which the objects are subsets of $L$ and morphisms represent inclusion of a smaller subset into a larger superset (i.e. $O \subseteq O' \Rightarrow O \rightarrow O'$).

Before proceeding, we define a few technical terms from the theory of sheaves and presheaves. We do not provide all necessary definitions to make use of the theory in more abstract contexts for which we direct the reader to~\cite{Lane1998}. Given $L$ we define a \textbf{presheaf} over it to be a contravariant functor, $PSh \colon\mathcal{P}(L)^{opp} \rightarrow \mathbf{Set}$, from the category of subsets of $L$, $\mathcal{P}(L)$, to the category of sets, $\mathbf{Set}$. Thus for every $U \in \mathcal{P}(L)$, $PSh(U)$ is a set.

$s \in PSh(U)$ is a \textbf{local section} over $U$ with respect to $PSh$. A \textbf{covering} of $L$ with respect to $\mathcal{P}(L)$ is an indexed set $\{O_i\}_{i \in I}$ where $O_i \subseteq L$ such that $\cup_{i \in I} O_i = L$.
A \textbf{system of local sections} over a covering $\{O_i\}_{i \in I}$ is a set of ordered pairs of elements, $O_i$, of the covering and sections, $s_i \in PSh(O_i)$, that comprise a set of the form
$$
\{(O_i,s_i) \; \vert \; i \in I, \; s_i \in PSh(O_i) \}.
$$
A system of local sections is \textbf{globally consistent} when there exists $s \in PSh(L)$ such that for all $i \in I$
$$
[ PSh( O_i \rightarrow L) ] (s) = s_i,
$$
where, $s$ is called a \textbf{witness} to global consistency.
A system of local sections is said to be \textbf{locally compatible} when for all $i,j \in I$ the following are satisfied
$$
[ PSh( O_{ij} \rightarrow O_i) ] (s_i) = [ PSh( O_{ij} \rightarrow O_j) ] (s_j),
$$
where $O_{ij} = O_i \cap O_j$.
Note that if a system of local sections is globally consistent then it is locally compatible. A presheaf is said to be a \textbf{sheaf} when given any locally compatible system of local sections, the system of local sections is both globablly consistent and there exists a unique witness. We refer to a presheaf that satisfies the existence but not the uniqueness condition as a \textbf{half-sheaf}.

The sheaf condition is also commonly expressed in terms of an equalizer diagram \cite{MacLane1992}. $PSh$ is a sheaf if beginning with the lattice of inclusions among subsets of network variables
\begin{equation}\label{eq:variablecoequalizer}
\xymatrix{
L \ar @{<--}[r]^-{\rho}
&
\coprod\limits_{i \in I}
O_{i}
\ar@{<-}@<1ex>[r]^-{\rho_i} \ar@{<-}@<-1ex>[r]_-{\rho_j}
&
\coprod\limits_{i,j \in I \times I} O_{ij}
},
\end{equation}
for any covering $\{O_i\}_{i \in I}$ and applying the $PSh$ functor to \autoref{eq:variablecoequalizer} results in
\begin{equation}\label{eq:sheafequalizer}
\xymatrix{
PSh(L) \ar @{-->}[r]^-{PSh(\rho)}
&
\coprod\limits_{i \in I}
PSh(O_i)
\ar@<1ex>[r]^-{PSh(\rho_i)} \ar@<-1ex>[r]_-{PSh(\rho_j)}
&
\coprod\limits_{i,j \in I \times I} PSh(O_{ij})
},
\end{equation}
where there exists $s \in PSh(L)$, such that all of the following conditions are satisfied
\begin{enumerate}
\item $[PSh(\rho)](s) = \{s|_{O_i} \; | \; i \in I\}$,
\item for a family $s_i \in PSh(O_i)$: $[PSh(\rho_i)] (s_i) = \{ s_i|_{O_{ij}} \}$ and $[PSh(\rho_j)] (s_i) = \{ s_j|_{O_{ij}} \}$,
\item $s$ is unique in satisfying conditions $1$ and $2$ among elements of $PSh(L)$.
\end{enumerate}
In this notation, if condition $3$ is not satisfied, then $PSh$ is a half-sheaf.

Given a presheaf and an associated covering, we may ask when it is the case that every locally compatible system of local sections over the covering is globally consistent. If this is the case, the covering is said to be \textbf{half-sheaf-like} because for the presheaves we study there is, in general, more than one witness to global consistency.

None of the presheaves we work with in this paper are sheaves, except in degenerate cases. We work exclusively with presheaves and their coverings. Some coverings are half-sheaf-like. Surprisingly, some are not. This is to say that, if a covering is not half-sheaf-like, then not every locally compatible system of local sections over the covering is globally consistent \cite{Boole1862,Vorobev1962,Bell1964,EthanAkin389,Abramsky2011}. The latter correspond to network architectures containing cycles whereas the former are acyclic.

A state of a subset of variables, $O \subseteq L$, is an assignment of values in $P$ to each variable in $O$ which is a tuple of length $|O|$ containing elements from $P$. This correspondence is determined by the presheaf functor $\expr = \mathrm{Hom}(-,P)$.  Specifically, this functor may be described as
\begin{equation}\label{eq:gpfunctor}
\begin{split}
\expr \colon \mathcal{P}(L)^{opp} &\to \mathbf{Set} \\
O &\mapsto P^O,\\
O \subseteq O' &\mapsto \{ e' \mapsto (e' \circ \iota) \mid e' \in P^{O'}\}, \end{split}
\end{equation}
where $\iota \colon O \rightarrow O'$ is the injection of the subset $O$ into $O'$ (i.e. $\iota (o) = o$ for all $o \in O$). In this case, $\expr$ is a sheaf, but note that this is not the case for the distribution presheaf, $\dist$, considered later.
For example, if we consider the case in which we have two variables $L=\{l_1,l_2\}$ and there are two potential states, $P=\{0,1\}$, then $\expr$ operates on the lattice of subsets generated by $L$ to give spaces of functions containing the possible \gnpm{} as exemplified in \autoref{fig:efunctor}.
For example, $\mathcal{E}(\{l_1,l_2\}) = \{ e^{12}_{00},e^{12}_{01},e^{12}_{10},e^{12}_{11} \}$ where $e^{12}_{01}(l_1) = 0$ and $e^{12}_{01}(l_2) = 1$. As another example, $\mathcal{E}(\{l_1\}) = \{ e^{1}_{0},e^{1}_{1} \}$ where $e^{1}_{0}(l_1)=0$ and $e^{1}_{1}(l_1)=1$. $\expr ( \{ l_1 \} \! \subseteq \! \{ l_1, l_2 \} )$ is given explicitly by
\begin{equation}
\begin{aligned}
e^{12}_{00} \mapsto e^{1}_{0}\\
e^{12}_{01} \mapsto e^{1}_{0}\\
e^{12}_{10} \mapsto e^{1}_{1}\\
e^{12}_{11} \mapsto e^{1}_{1}
\end{aligned}.
\end{equation}

Next, we introduce extended probability distributions by defining a functor $\mathcal{D}$ that will compose with $\mathcal{E}$ to convert collections of \gnpm{} into probability distributions over them.  Given a finite set $S$, define $\dist(S)$ to be the set of all maps from $\mathcal{P}(S)$ to the interval $[0,1]$ which satisfy the following two conditions:  For all $d \in \dist(S)$, we have $d(S) \in \{0,1\}$.\footnote{Normally, we would only have $d(S) = 1$, but since we want to introduce conditionalization in a coherent way it becomes necessary to admit degenerate distributions where $d(S) = 0$ as well. This simplifies the exposition by not requiring us to worry about dividing by zero and having to introduce special cases when dealing with conditional probabilities and partial functions.  Of course, it also means that we cannot automatically assume that an element of $\dist(S)$ can be normalized without checking this fact but in our examples, this verification will turn out to be routine and trivial.}  For all  $d \in \dist(S)$ and all $A, B \subset S$, we have $d(A) + d(B) = d(A \cup B) + d(A \cap B)$.

Returning to the running example,
\begin{equation}\label{eq:exampledistfromfunctors}
\begin{aligned}
\mathcal{D}(\mathcal{E}(\{l_1,l_2\})) &= \{p^{12}_{00},p^{12}_{01},p^{12}_{10},p^{12}_{11} \mid p^{12}_{00} \geq 0, p^{12}_{01} \geq 0,p^{12}_{10} \geq 0,p^{12}_{11} \geq 0, p^{12}_{00} + p^{12}_{01} + p^{12}_{10} + p^{12}_{11} = 1 \},\\
\mathcal{D}(\mathcal{E}(\{l_1\})) &= \{p^{1}_{0}, p^{1}_{1} \mid p^{1}_{0} \geq 0, p^{1}_{1} \geq 0, p^{1}_{0}+p^{1}_{1} = 1 \}.
\end{aligned}
\end{equation}

If $S$ and $S'$ are finite sets, which in our case will usually be sets of \gnpm{} given by $\mathcal{E}(O)$, $d \in \dist(S)$ and $d' \in \dist(S')$ are probability distributions over these spaces, and $f \colon S \to S'$ is a partial function, we will say that $f$ is compatible with $d$ and $d'$ when, for all $X \in \mathrm{img}(f)$, we have
\begin{equation}
 d'(X) =
  \begin{cases}
    \frac{d'(\mathrm{img}(f))}{d(\mathrm{dom}(f))}d(f^{-1}(X)) & d(\mathrm{dom}(f)) \neq 0, \\
    0 & d(\mathrm{dom}(f)) = 0.
  \end{cases}
\end{equation}
In other words, the map $f$ preserves ratios of probabilities of events.  In the case where $f$ is a partial surjection $(\mathrm{img}(f)=S')$, compatibility completely determines $d'$ in terms of $d$ and thus $\dist$ may be regarded as a functor from the subcategory of sets with partial surjections as morphisms to transformations on probability distributions:
\begin{equation}\label{eq:distfunctor}
 d' = \dist(f)(d) =
  \begin{cases}
    X \mapsto \frac{d(f^{-1}(X))}{d(\mathrm{dom}(f))} & d(\mathrm{dom}(f)) \neq 0, \\
    X \mapsto 0 & d(\mathrm{dom}(f)) = 0.
  \end{cases}
\end{equation}
Specifically, when $f$ is a total surjection, this map corresponds to marginalization. For example, in the case $f = \expr ( \{ l_1 \} \! \subseteq \! \{ l_1, l_2 \} )$
\begin{equation}\label{eq:dfunctorex}
\begin{split}
\dist \big(\expr ( \{ l_1 \} \! \subseteq \! \{ l_1, l_2 \} ) \big) \colon \expr ( \{ l_1 \} \! \subseteq \! \{ l_1, l_2 \} ) &\to \expr ( \{ l_1 \} ), \\
d &\mapsto d',
\end{split}
\end{equation}
then
\begin{equation}\label{eq:margexample}
d\hspace{.09em}'(\{e^1_0\}) = \dist (f)(d)(\{e^1_0\}) = \frac{d(f^{-1}(\{ e^1_0\}))}{d(\{ e^{12}_{00},e^{12}_{01},e^{12}_{10},e^{12}_{11} \})} = \frac{d(\{ e^{12}_{00},e^{12}_{01}\})}{1} = d( \{ e^{12}_{00} \} ) + d( \{ e^{12}_{01} \} ) = p^{12}_{00} + p^{12}_{01}.
\end{equation}
When $f$ is a partial isomorphism, it corresponds to conditionalization. For example, if $f$ is defined such that we condition on variable one being in state zero, $l_1 = 0$,
\begin{equation}
f \colon \{ e^{12}_{00} , e^{12}_{01} \} \subset \expr ( \{ l_1, l_2 \} ) \to \{ f(e^{12}_{00}) , f(e^{12}_{01}) \}
\end{equation}
then
\begin{equation}\label{eq:condexample}
d\hspace{.09em}'(\{ f(e^{12}_{00}) \} ) = \dist (f)(d)( \{ f(e^{12}_{00}) \}) = \frac{d( f^{-1}( \{ f(e^{12}_{00}) \}) )}{d(\{ e^{12}_{00},e^{12}_{01} \})} = \frac{d(\{ e^{12}_{00} \})}{d( \{ e^{12}_{00} \} ) + d( \{ e^{12}_{01} \} )} = \frac{p^{12}_{00}}{p^{12}_{00} + p^{12}_{01}}.
\end{equation}
Finally, when $f$ is a general partial surjection, it corresponds to a combination of conditionalization and marginalization.

In order to admit the basic tools of linear algebra for the purpose of calculations regarding relationships between spaces of probability distributions we explain how they embed into linear spaces. By definition, an extended probability distribution $p \in \dist(S)$ is an element of $\mathbb{R}^S$. We denote the inclusion map as
\begin{equation}\label{eq:emb}
\rm{emb}(S) : \dist(S) \to \mathbb{R}^S.
\end{equation}
Because a convex combination of two probability distributions is again a probability distribution, the image of $\mathrm{emb}(S)$ consists of a convex set and the origin point (corresponding to the degenerate zero distribution).  Furthermore, if $n$ is the number of elements of the set $S$, this convex set works out to be the probability simplex with $n$ vertices, which we denote $\Delta_{n-1}$.  In our example above, $\mathcal{D}(\mathcal{E}(\{l_1,l_2\}))$ is the tetrahedron $\Delta_3$. Since any vector $v \in \mathbb{R}^S$ may be written as $v = c_{+} p_{+} - c_{-} p_{-}$ where $c_{+}, c_{-} \in [0,\infty)$ and $p_{+}$ and $p_{-}$ are probability distributions, the image of $\mathrm{emb}(S)$ spans the vector space $\mathbb{R}^S$.  For purposes of later reference, note that, if $f \colon S \to S'$ is a partial surjection, then $\mathcal{D}$ extends to a fractional linear map, as in \autoref{eq:condexample}, from $\mathbb{R}^S$ to $\mathbb{R}^{S'}$ and that, in the special case where $f$ is a total surjection, as in \autoref{eq:margexample}, it is in fact a linear map.

\section{Precise formulation of coarse-graining network states}\label{secsupp:coarsegrainingphenotypes}

As described in \autoref{sec:genenetworkphenmap} it is also possible to consider network states that derive from coarse-graining lower-level network states. Once this is done, one arrives at probability distributions over network modules like that introduced in \autoref{sec:probabilitydistributionsonnetworks}. As a result of this, our conclusions that are formulated in terms of a single level of coarse-graining \gnpm{} also apply to coarse-graining over multiple levels at once despite the fact that the parameters of the relevant probabilistic model are likely to be different.

For each subset of variables $O \in \mathcal{P}(L)$, let $\phi_i (O)$ be the set of network states at level $i$, which can be determined from the expression levels of variables in $O$.  Note that $\phi_i (O)$ may be empty if the set $O$ does not contain enough variables to determine the values of any network state at level $i$. When $O_1 \subseteq O_2 \in \mathcal{P}(L)$, we have a restriction map $\pi_i^{O_2 O_1} \colon \phi_i(O_2) \to \phi_i(O_1)$.  These maps satisfy the consistency conditions that $\pi_i^{OO}$ is the identity map and that $\pi_i^{O_3 O_2} \circ \pi_i^{O_2 O_1} = \pi_i^{O_3 O_1}$, i.e. $\pi_i$ is a functor on $(\mathcal{P}(L), \subseteq)$.  As stated earlier, we set $\phi_1 (O) = P^O$ and $\pi_1^{O_2 O_1}$ to be the restriction map from $P^{O_2}$ to $P^{O_1}$.  If $i \le j$, let $\Omega_{ij}(O) : \phi_i(O) \to \phi_j(O)$ be the coarse-graining map which describes how higher level network states are determined from lower level network states.  These maps are all surjections and, for consistency, we will require the following conditions:
\begin{enumerate}
\item $\Omega_{ij}(O) \circ \Omega_{jk}(O) = \Omega_{ik}(O)$ whenever $i \le j \le k$.
\item $\Omega_{ii}(O)$ is the identity map on $\phi_i (O)$.
\item If $O_1 \subseteq O_2 \in \mathcal{P}(L)$ and $i > j$, then $\Omega_{ij}(O_1) \circ \pi_i^{O_2 O_1} = \pi_j^{O_2 O_1} \circ \Omega_{ij}(O_2)$
\end{enumerate}
In other words, $\Omega$ must be suitably functorial in both of its arguments.

\begin{figure}[!ht]
\centering
\noindent\includegraphics[width=0.9\columnwidth]{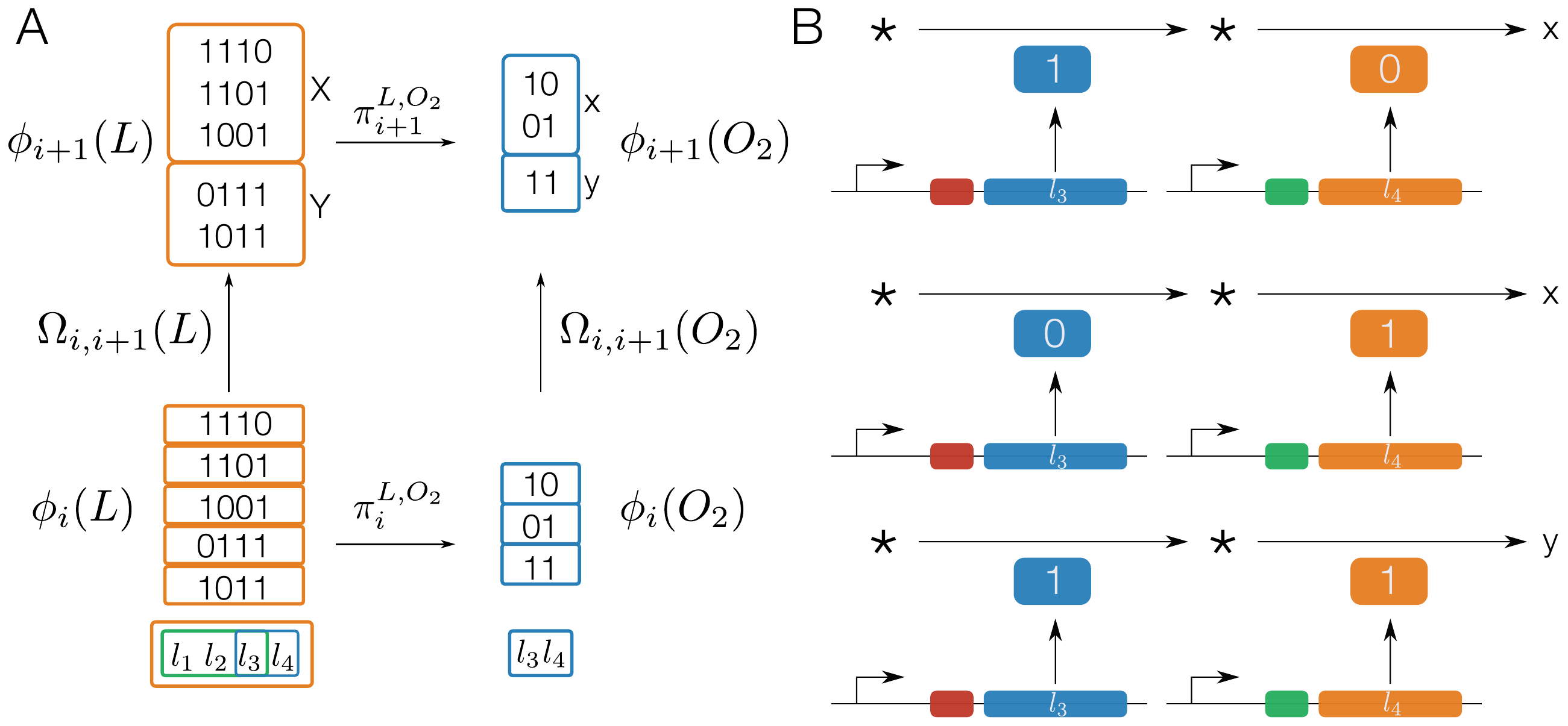}
\caption{{\bf Example coarse-graining of phenotypes.} (A) Consider the example where $L = \{l_1,l_2,l_3,l_4\}$, $\mathcal{G} = \{ O_1, O_2 \}$, $O_1 = \{l_1, l_2, l_3\}$ and $O_2 = \{ l_3, l_4 \}$. The top left panel shows two higher-level phenotypes $X$ and $Y$. The bottom left corner shows the five different expression states of four genes in $L$ from which these phenotypes are coarse-grained. The right side shows the respective projections onto genes $\{l_3,l_4\}$. The projection maps $\pi_i^{L,O_2}$ and $\pi_{i+1}^{L,O_2}$ are defined in \refsupp{} \autoref{secsupp:coarsegrainingphenotypes}. (B) The different combinations of expression states of genes $\{l_3,l_4\}$ result in two different phenotypes. If both genes are expressed metabolite $y$ is produced whereas if only one of the two genes is expressed metabolite $x$ is produced. The red and green boxes represent arbitrary promoters.}
\label{fig:phenotypehierarchy}
\end{figure}

For example, if our lower level network states for a set of variables $O_1 = \{ l_1,l_2,l_3,l_4 \}$ are given by a set of binary sequences, then the projection of these network states down to the set $O_2 = \{l_3,l_4\}$ followed by mapping to the higher level network states $x=\{01,10\}$ and $y=\{11\}$ is equivalent to first mapping to the higher-level network states $X$ and $Y$ and then projecting down to $O_2$ shown by the equivalent paths from the top-left to the bottom-right in \autoref{fig:phenotypehierarchy}A.
Of course, there is an equivalent diagram for the subset $\{ l_1,l_2,l_3 \}$.

\begin{figure}[!ht]
\centering
\noindent\includegraphics[width=0.4\columnwidth]{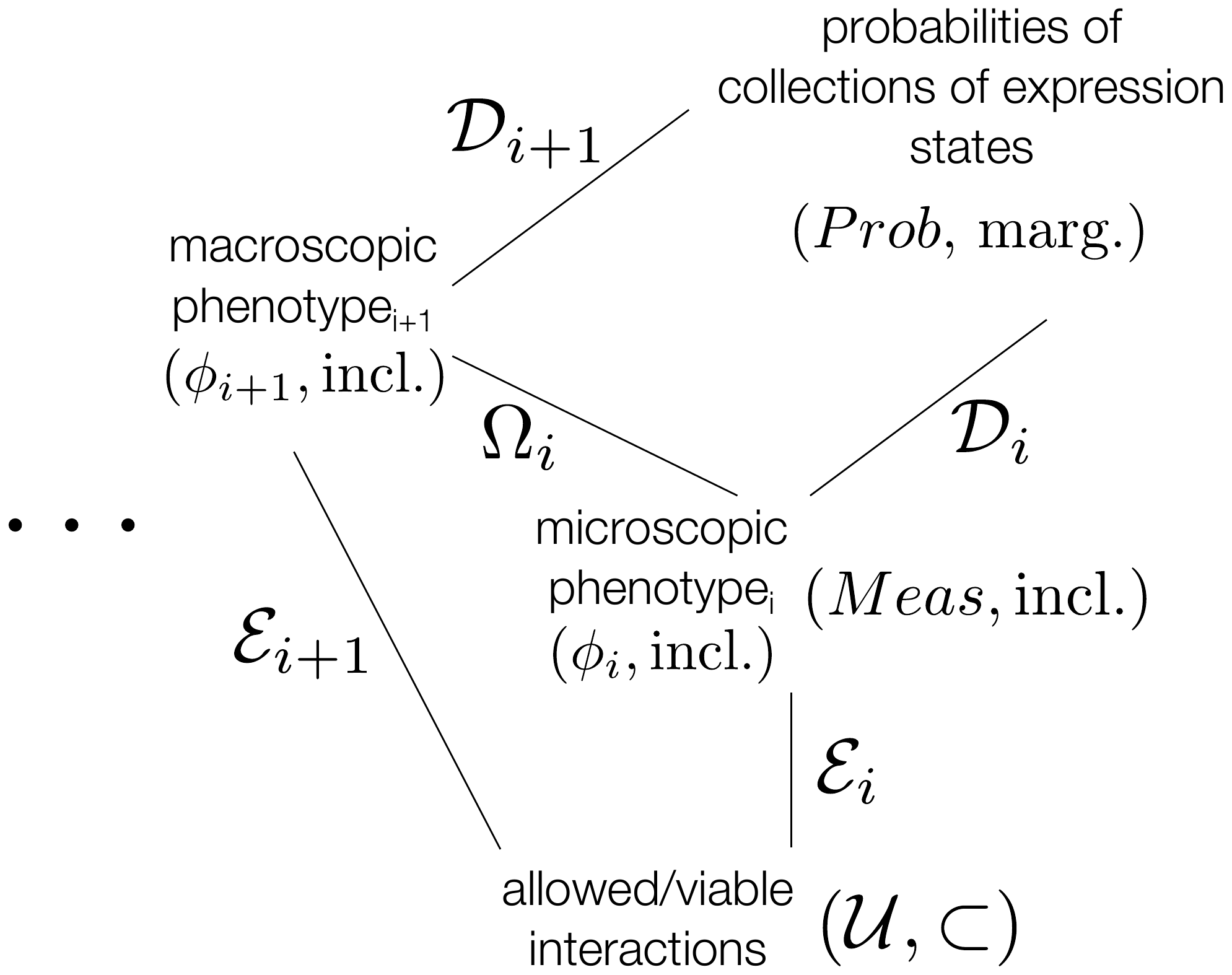}
\caption{{\bf Mathematical relationships defining the hierarchy of network states via coarse-graining.} }
\label{fig:abstractroadmap}
\end{figure}

Since the map $\Omega_{1i}(O)$ is a surjection from $P^O$ onto $\phi_i(O)$, we can use it to map our probabilistic structures to $\phi_i(O)$.  Set $\mathcal{E}_i = \Omega_{1i}(O)^{-1} \circ \mathcal{E}$ and $\dist_i = \Omega_{1i}(O)^{-1} \circ \dist$.  Then we end up with the overall relationships summarized in \autoref{fig:abstractroadmap}. As a consequence of the consistency conditions the coarse-graining maps $\phi$ and $\Omega$, there is a natural transformation between the functors $\expr_i$ and $\expr_{i+1}$ implying that the following diagram commutes
$$
\xymatrix{
\expr_{i+1}(O_1) \ar[r]^{\expr_{i+1}(\subseteq)} \ar[d]_{t_{O_1}} & \expr_{i+1}(O_2) \ar[d]^{t_{O_2}} \\
\expr_{i}(O_1) \ar[r]^{\expr_{i}(\subseteq)} & \expr_{i}(O_2) }
$$
for any $O_2 \subseteq O_1$.

Given a covering $\mathcal{G}$ of the space of biological network variables, we can consider the higher order network states associated to the elements of $\mathcal{G}$.  For a suitable choice of cover and a suitable level of network states, it may happen that the network states associated to different elements of $\mathcal{G}$ are distinct.  For instance, in the example of \autoref{fig:phenotypehierarchy}, if we take $\mathcal{G} = \{O_1, O_2\}$ where $O_1 = \{l_1, l_2, l_3\}$ and $O_2 = \{l_3, l_4\}$, we have $\phi_{i+1}(O_1) = \{u,v\}$ and $\phi_{i+1}(O_2) = \{x,y\}$.  In such a case, if we were to perform one experiment which measured the network states $\{u,v\}$ and another experiment which measured $\{x,y\}$, then the result could be understood as examining the covering $\{O_1, O_2\}$ at network state level $i+1$.

\section{Sheaf-theoretic formulation of compatibility of distributions on \gnpm{}}\label{secsupp:compatibilityofgpms}

Given a covering of the space of variables $\mathcal{G}$, a compatible family for $\mathcal{G}$ with respect to $\mathcal{D} \circ \mathcal{E}$ is given by a family of distributions $\dist{}(\expr{}(\mathcal{G})) = \{d_O \in \dist (\mathcal{E}(O)) | O \in \mathcal{G}\}$ such that for all $O, O' \in \mathcal{G}$
\begin{eqnarray}\label{eq:sheafcond}
d_O|O \cap O' = d_{O'}|O \cap O'.
\end{eqnarray}
This first set of conditions is later referred to as local consistency. The space of all such locally consistent distributions for a given covering, $\mathcal{G}$, is referred to as $\mathbb{L}(\mathcal{G})$ where
\begin{equation}
\mathbb{L}(\mathcal{G}) = \{ d_O  \in \dist{}(\expr{}(\mathcal{G})) \mid (\forall O,O' \in \mathcal{G})\,\, d_O|O \cap O' = d_O'|O \cap O' \}.
\end{equation}
These conditions mean that any two distributions $d_O$ and $d_{O'}$ in the \emph{compatible family} of distributions marginalize to the same disribution over the intersection of $O$ with $O'$. If these constraints are not satisfied, then there is no way to make a consistent assignment of probabilities to the states of even a single variable. In this case in order to restore consistency one of the constraints must be eliminated or duplication of a variable may allow for the independent satisfaction of both constraints.

If, moreover, this first condition implies the existence of $d \in \mathcal{D}( \mathcal{E}(L))$ such that $d|O = d_O$ for all $O \in \mathcal{G}$ then the system is said to satisfy the global consistency condition.
The space of all such globally consistent distributions for a given covering, $\mathcal{G}$, is referred to as $\mathbb{M}(\mathcal{G})$ where
\begin{equation}
\mathbb{M}(\mathcal{G}) = \{ d_O \in \dist{}(\expr{}(\mathcal{G})) \mid (\exists d ) \,\, d|O = d_O \}.
\end{equation}
In general, the system of equations $d|O = d_O$ for all $O \in \mathcal{G}$ is underdetermined and so local consistency does not imply global consistency.  Local and global consistency are formalized as described in \ref{secsupp:probabilitydistributionsonnetworks} in terms of sheaf theory as applied to the presheaf functors $\expr$ and $\dist \circ \expr$. $\mathcal{E}$ alone turns out to be a sheaf because it satisfies the analogous conditions for all possible coverings $\mathcal{G}$ of $L$: for $\{e_O \in \mathcal{E}(O) | O \in \mathcal{G}\}$ such that $e_{O_{1}} | O_1 \cap O_2 = e_{O_{2}} | O_1 \cap O_2$ there exists a unique $e \in \mathcal{E}(\cup_{O \in \mathcal{G}} O)$ such that $e_O = e|O$ for all $O \in \mathcal{G}$. By analogy to \autoref{eq:sheafequalizer} this is expressed by applying the same conditions to the equalizer diagram
\begin{equation}\label{eq:tuplesheaf}
\xymatrix{
\expr(L) \ar @{-->}[r]^-{e}
&
\coprod\limits_{i \in I}
\expr(O_i)
\ar@<1ex>[r]^-{e_{O_i}} \ar@<-1ex>[r]_-{e_{O_j}}
&
\coprod\limits_{i,j \in I \times I} \expr(O_{ij})
}.
\end{equation}
For $\dist \circ \mathcal{E}$ the sheaf condition is not automatically satisfied and it only defines a presheaf. We examine the situation more closely to explicitly determine the necessary conditions for global consistency.

For a cover of the space of variables, $\mathcal{G}$, we can construct a linear operator, $\mathbf{G}$, representing the relationship, $R = \coprod_{O \in \mathcal{G}} \mathcal{E}(O \subset L) \subseteq \mathcal{E}(L) \times \expr (\mathcal{G})$, between \gnpm{} having as domain particular network modules given by the $O \in \mathcal{G}$ and those global \gnpm{} defined on $L$. We would like to construct the matrix representation of $\mathbf{G}$. In the first factor, $\mathcal{E}(L) = P^L = \{e^{L}_{\vec{j}} | \vec{j} \in P^{|L|}\}$.
For the second factor, $ \mathcal{E}(\mathcal{G}) = \coprod_{O \in \mathcal{G}} \mathcal{E}(O) = \{e^O_{\vec{i}} | O \in \mathcal{G}, \vec{i} \in P^{|O|} \}$.
 So we have two sets of maps, one defined on $P^L$ and the other defined on $\mathcal{E}(O) = P^O$ for each $O \in \mathcal{G}$. This yields the method of specifying the intended relationship that defines $\mathbf{G}$ for all $e^O_{\vec{i}} \in \mathcal{E}(\mathcal{G})$ and $e^{L}_{\vec{j}} \in \mathcal{E}(L)$ given in \autoref{eq:margmat}.
This matrix can be viewed as an operator acting via matrix multiplication on distributions
\begin{eqnarray*}
\mathbf{G} \colon \mathcal{D}(\mathcal{E}(L)) &\rightarrow& \mathcal{D}( \mathcal{E}(\mathcal{G})),\\
d &\mapsto& \coprod_{O \in \mathcal{G}} d|O,
\end{eqnarray*}
and thereby taking a global distribution, $\dist (\mathcal{E}(L))$, defined on \gnpm{} whose domain is the full set of variables $L$ into the local distributions, $\dist (\mathcal{E}(\mathcal{G}))$, that are defined relative to network modules contained in a covering of the space of variables $\mathcal{G}$.
$\mathbf{G}$ can be specified for all $e^O_{\vec{i}} \in \mathcal{E}(\mathcal{G})$ and $e^{L}_{\vec{j}} \in \mathcal{E}(L)$:
\begin{eqnarray}\label{eq:margmat}
\mathbf{G}(e^O_{\vec{i}},e^L_{\vec{j}}) =
\begin{cases}
1, & e^L_{\vec{j}}|O = e^O_{\vec{i}},\\
0, & \text{otherwise}.
\end{cases}
\end{eqnarray}
For example, given the covering $\mathcal{G} = \{ \{l_1\}, \{l_2\} \}$ of a set of two variables $L = \{ l_1, l_2 \}$ the associated matrix $\mathbf{G}$ is shown in \autoref{fig:efunctor}B.
$\mathbf{G}$ provides a way of determining the distributions on \gnpm{} for a given context (i.e. $\coprod_{O \in \mathcal{G}} d|O$) that can be derived from distributions (i.e. $\mathcal{D} ( \mathcal{E}(L) )$) defined on the global \gnpm{} (i.e. $\mathcal{E}(L)$ as opposed to $\mathcal{E}(O)$).

Having expressed the relationship between global and local network-network state maps in terms of $\mathbf{G}$ we now make use of sheaf theory in order to extract the global consistency conditions. Given \autoref{eq:tuplesheaf} and the associated conditions making $\expr$ a sheaf, $\mathbb{R}^{\expr}$ given by
\begin{equation}\label{eq:realtuplesheaf}
\xymatrix{
\mathbb{R}^{\expr(L)} \ar @{->}[r]^-{\mathbf{G}}
&
\bigoplus\limits_{i \in I}
\mathbb{R}^{\expr(O_i)}
\ar@<1ex>[r]^-{\mathbf{H_1}} \ar@<-1ex>[r]_-{\mathbf{H_2}}
&
\bigoplus\limits_{i,j \in I \times I} \mathbb{R}^{\expr(O_{ij})}
},
\end{equation}
is a half-sheaf, in the sense that it satisfies the first two conditions but not the third uniqueness condition given in \autoref{secsupp:probabilitydistributionsonnetworks}. It follows from this fact that $ker(\mathbf{H}_1 - \mathbf{H}_2) = im(\mathbf{G})$.  Moreover, although $\dist \circ \expr$ is a mere presheaf, it can be embedded into $\mathbb{R}^{\expr}$ using the map defined in \autoref{eq:emb} thereby allowing for the expression of consistency conditions on $\dist \circ \expr$ in terms of linear equations constituting constraints on the relevant probabilities.
The following diagram demonstrates the relationships between the spaces of probability distributions and the linear spaces in which they are embedded:
\begin{equation}\label{eq:embeddingdiagram}
\xymatrix{
 \mathbb{R}^{\mathcal{E}(L)} \ar[r]^{\mathbf{G}} &
   \mathbb{R}^{\mathcal{E}(\mathcal{G})} \\
 \dist (\mathcal{E}(L)) \ar[r]^{\mathbf{G}} \ar@{^{(}->}[u]^{emb_{\mathcal{E}(L)}}& \ar@{^{(}->}[u]_{emb_{\mathcal{E}(\mathcal{G})}}
  \dist (\mathcal{E}(\mathcal{G}))
  }
\end{equation}
The locally, $\mathbb{L}(\mathcal{G})$, and globally, $\mathbb{M}(\mathcal{G})$, consistent polytopes correspond to the spaces of probability distributions satisfying the local and global consistency conditions described above.
In terms of the diagrams expressing the half-sheaf condition, \autoref{eq:realtuplesheaf}, and embedding map, \autoref{eq:embeddingdiagram},
\begin{equation}
\begin{aligned}
\mathbb{M}(\mathcal{G}) &= \mathbf{G}(emb_{\mathcal{E}(L)}(\mathcal{D}(\mathcal{E}(L))))\\
\mathbb{L}(\mathcal{G}) &= emb_{\mathcal{E}(\mathcal{G})}(\mathcal{D}(\mathcal{E}(\mathcal{G}))) \cap ker(\mathbf{H}_1(\mathbb{R}^{\mathcal{E}(\mathcal{G})})-\mathbf{H}_2(\mathbb{R}^{\mathcal{E}(\mathcal{G})})) = emb_{\mathcal{E}(\mathcal{G})}(\mathcal{D}(\mathcal{E}(\mathcal{G})))\cap\mathbf{G}(\mathbb{R}^{\mathcal{E}(L)}).
\end{aligned}
\end{equation}
As in \autoref{eq:exampledistfromfunctors}
\begin{eqnarray}
emb_{\mathcal{E}(\mathcal{G})}(\mathcal{D}(\mathcal{E}(\mathcal{G}))) &=& \left\{ p^{O}_{\vec{i}} \; \bigg| \; (\forall\, O \in \mathcal{G}) \; (\forall\, \vec{i} \in P^{|O|}) \;\; p^O_{\vec{i}} \geq 0, \; (\forall\, O \in \mathcal{G}) \sum_{\vec{i} \in P^{|O|}} p^{O}_{\vec{i}} = 1 \right\}, \label{eq:embdeg}\\
emb_{\mathcal{E}(L)}(\mathcal{D}(\mathcal{E}(L))) &=& \left\{ p^L_{\vec{i}} \; \bigg| \; (\forall\, \vec{i} \in P^{|L|}) \; p^L_{\vec{i}} \geq 0, \; \sum_{\vec{i} \in P^{|L|}} p^L_{\vec{i}} = 1 \right\}. \label{eq:embdel}
\end{eqnarray}
In general the globally consistent polytope is a proper subspace of the locally consistent one because $\mathbf{G}$ is not invertible (the maximum entropy principle is commonly used to make an arbitrary choice in the face of this underdetermination).
To determine explicit conditions on the probabilities we express $\mathbb{L}(\mathcal{G})$ and $\mathbb{M}(\mathcal{G})$ in terms of the fundamental subspaces associated to the linear map $\mathbf{G}$. In order for a vector $v$ to lie in $\mathbf{G}(\mathbb{R}^{\mathcal{E}(L)})$, we must have $v=\mathbf{G}x$ for some $x \in \mathbb{R}^{\mathcal{E}(L)}$.  The cokernel of $\mathbf{G}$ gives the obstructions to this system $v=\mathbf{G}x$ having a solution. In order to eliminate these obstructions, constraints must be imposed on $\mathbb{R}^{\mathcal{E}(\mathcal{G})}$ and these constraints are given precisely via annihilating the cokernel, i.e. $\mathbf{G}(\mathbb{R}^{\mathcal{E}(L)}) = \{ v \; | \; (\forall u \in coker \mathbf{G}) \; u \cdot v = 0\}$.
We then take the appropriate intersection to determine $\mathbb{L}(\mathcal{G})$ by requiring $v \in \mathcal{D}(\mathcal{E}(\mathcal{G}))$
\begin{eqnarray}\label{eq:localpolytope}
\mathbb{L}(\mathcal{G}) &=& \{ v \in \dist{}(\expr{}(\mathcal{G})) \; | \; (\forall u \in coker \mathbf{G}) \; u \cdot v = 0\}.
\end{eqnarray}
Since $\mathbf{G}$ is not invertible the equation $v = \mathbf{G}x$ can only be solved up to an element of $ker \mathbf{G}$. $v = \mathbf{G}x$ can thus be solved on a subspace $T$ of $\mathbb{R}^{\mathcal{E}(L)}$ such that $T \oplus ker \mathbf{G} = \mathbb{R}^{\mathcal{E}(L)}$ to yield
\begin{eqnarray}\label{eq:globalpolytope}
\mathbb{M}(\mathcal{G}) &=& \{ v \; | v = \mathbf{G} x,\; (\exists x \in T) \; (\exists y \in \dist{}(\expr{}(L)))\; x-y \in ker \mathbf{G}  \}.
\end{eqnarray}
If the embedding into linear spaces is to be considered explicitly, then \autoref{eq:embdeg} and \autoref{eq:embdel} can be substituted for $\dist{}(\expr{}(\mathcal{G}))$ and $\dist{}(\expr{}(L))$ in \autoref{eq:localpolytope} and \autoref{eq:globalpolytope}.
In order to obtain inequalities that define $\mathbb{M}(\mathcal{G})$, Fourier-Motzkin elimination can be used to eliminate $x$ and $y$. Alternatively one can use the fact, \cite{Wainwright2007} proposition 8.3, that $\mathbb{M}(\mathcal{G})$ is given by removing the non-integer vertices from a vertex representation of $\mathbb{L}(\mathcal{G})$ and the ability to interconvert between vertex and inequality representations to compute the same inequalities as described in \refsupp{} \autoref{secsupp:fourcycleexample}.

\subsection{Example of apparent satisfaction of unsatisfiable constraints}\label{secsupp:apparentinconsistency}
The inequalities defining $\mathbb{M}(\mathcal{G})$ were derived under the assumption that the two-element probabilities were obtained by mariginalizing a three-element distribution.  If some other procedure, such as conditionalization, is used to obtain them instead, these inequalities need not apply. For example, suppose now that $L = \{l_1,l_2,l_3 \}$, $P = \{0,1,2\}$, $\mathcal{G} = \{\{l_1,l_2\},\{l_2,l_3\},\{l_3,l_1\}\}$ where we have simply added an element to $P$ relative to the example described above. In the previous example the marginal maps were given by $\dist (\expr (O \subset L))$ with one for each $O \in \mathcal{G}$. If we combine these marginal maps with conditioning on one out of the three variables being in state two and each of the other two being in states zero or one, then we have instead $\dist (\pi_1), \dist (\pi_2), \dist (\pi_3)$ where $\pi_1 = \expr (\{l_1,l_2\} \subset L)|\{ e^{123}_{ij2} \mid i,j \in \{ 0,1 \} \},\; \pi_2 = \expr (\{l_2,l_3\} \subset L)|\{ e^{123}_{2ij} \mid i,j \in \{ 0,1 \} \},\; \pi_3 = \expr (\{l_3,l_1\} \subset L)|\{ e^{123}_{i2j} \mid i,j \in \{ 0,1 \} \}$. In this case, if we have the following assignment of probabilities for a distribution $d$
\begin{equation}\label{eq:condprobs}
\begin{aligned}
p^{123}_{002} &= 1/30 & p^{123}_{020} &= 2/15 & p^{123}_{200} &= 2/15\\
p^{123}_{012} &= 2/15 & p^{123}_{021} &= 1/30 & p^{123}_{201} &= 1/30\\
p^{123}_{102} &= 2/15 & p^{123}_{120} &= 1/30 & p^{123}_{210} &= 1/30\\
p^{123}_{112} &= 1/30 & p^{123}_{121} &= 2/15 & p^{123}_{211} &= 2/15
\end{aligned}
\end{equation}
with all other probabilities being zero, then $\dist (\pi_1)(d), \dist (\pi_2)(d), \dist (\pi_3)(d)$ are equivalent to the probability tables in \autoref{fig:inconsistentthreecycle}A, which as shown in \autoref{sec:inconsistency}, could not be achieved by marginalization alone. For example, given that $dom(\pi_1) = \{ e^{123}_{002}, e^{123}_{012}, e^{123}_{102}, e^{123}_{112} \}$ then $d(dom(\pi_1)) = \frac{1}{30} + \frac{2}{30} + \frac{2}{30} + \frac{1}{30} = \frac{1}{3}$. Substituting this factor and the fact that ${\pi_1}^{-1}(e^{12}_{ij}) = e^{123}_{ij2}$ into \autoref{eq:distfunctor}
$$
p^{12}_{ij} = \dist (\pi_1)(d)(e^{12}_{ij}) = \frac{d({\pi_1}^{-1}(e^{12}_{ij}))}{d(dom(\pi_1))} = 3 p^{123}_{ij2},
$$
then renormalizes probabilities resulting in $p^{12}_{00} = 0.1, p^{12}_{01} = 0.4, p^{12}_{10} = 0.4, p^{12}_{11} = 0.1$ along with the analogs for $p^{23}_{ij}$ and $p^{13}_{ij}$, which are precisely equivalent to what appears in \autoref{fig:inconsistentthreecycle}A as suggested above.

If constraints consistent with those of \autoref{fig:inconsistentthreecycle}A are placed on the given network, either the network must add another variable in order to satisfy them directly or the network context imposing those constraints must coarse-grain the network in a suitable way. In what follows, we argue that the former is much more plausible than the latter. This ultimately suggests conditions in which cycle breakage may be selected for to relieve inconsistent constraints that can arise when cycles are present.

\section{Example volume ratio computation for the four-cycle network architecture}\label{secsupp:fourcycleexample}

\begin{figure}[!ht]
\centering
\noindent\includegraphics[width=0.8\columnwidth]{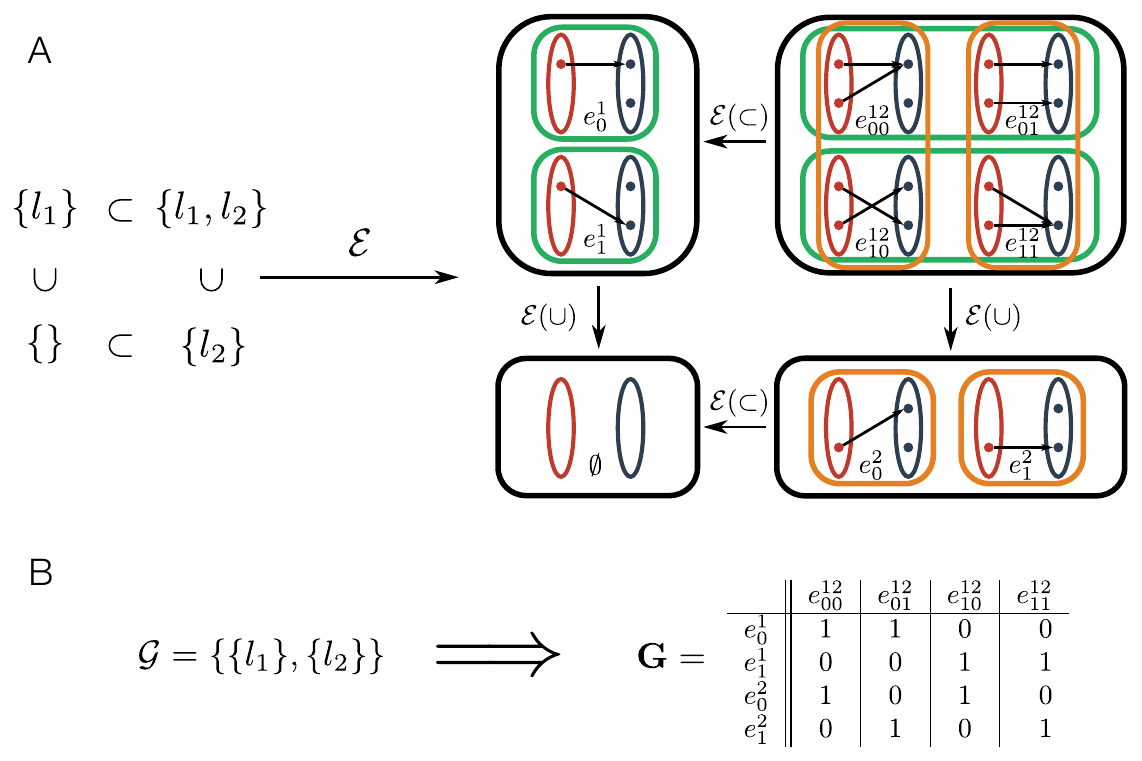}
\caption{{\bf Example of the functor mapping subsets of variables to measurable spaces.} (A) On the left hand side are subsets of $L=\{l_1,l_2\}$ ordered by inclusion. On the right hand side are the spaces of \gnpm{} also ordered by inclusion. The labels for the maps define them. For example, $e^{12}_{01}(l_1) = 0$ and $e^{12}_{01}(l_2) = 1$. (B) For the given covering, $\mathcal{G}$, the associated marginalization matrix acting on the probability vector $\{ p^{12}_{00},p^{12}_{01},p^{12}_{10},p^{12}_{11} \}$ to give $\{ p^{1}_{0},p^{1}_{1},p^{2}_{0},p^{2}_{1} \}$ is $\mathbf{G}$.}
\label{fig:efunctor}
\end{figure}

For the purposes of this example, we take the full set of variables to be $L = \{ l_1,l_2,l_3,l_4 \}$. Consider the case in which each of the network modules under consideration has two variables and we specify the covering of the space of variables given by $\mathcal{G} = \{\{l_1,l_2 \},\{l_1,l_4 \},\{l_3,l_2\},\{l_3,l_4\} \}$.  We will compute $\mathbb{L}(\mathcal{G})$ using the same method which was used for the example of three variables.  By analogy with \autoref{eq:localconsistencythreegenes}, the local consistency conditions now are as follows:
\begin{equation}
\begin{aligned}\label{eq:pparsys}
p^1_0 &= p^{12}_{00} + p^{12}_{01} = p^{14}_{00} + p^{14}_{01}, &
p^1_1 &= p^{12}_{10} + p^{12}_{11} = p^{14}_{10} + p^{14}_{11},\\
p^3_0 &= p^{32}_{00} + p^{32}_{01} = p^{34}_{00} + p^{34}_{01},&
p^3_1 &= p^{32}_{10} + p^{32}_{11} = p^{34}_{11} + p^{34}_{11},\\
p^2_0 &= p^{12}_{00} + p^{12}_{10} = p^{32}_{00} + p^{32}_{10},&
p^2_1 &= p^{12}_{01} + p^{12}_{11} = p^{32}_{01} + p^{32}_{11},\\
p^4_0 &= p^{14}_{00} + p^{14}_{10} = p^{34}_{00} + p^{34}_{10},&
p^4_1 &= p^{14}_{01} + p^{14}_{11} = p^{34}_{01} + p^{34}_{11}.
\end{aligned}
\end{equation}
Likewise, the equations determined by the conditions $v = \mathbf{G}(x)$ which are analogous to the matrix $\mathbf{G}$ in \autoref{fig:efunctor}B are now
\begin{equation}
\begin{aligned}\label{eq:globsys}
p^{12}_{00} &= p^{1234}_{0000} + p^{1234}_{0010} + p^{1234}_{0001} + p^{1234}_{0011} &
p^{12}_{10} &= p^{1234}_{1000} + p^{1234}_{1010} + p^{1234}_{1001} + p^{1234}_{1011} \\
p^{12}_{01} &= p^{1234}_{0100} + p^{1234}_{0110} + p^{1234}_{0101} + p^{1234}_{0111} &
p^{12}_{11} &= p^{1234}_{1100} + p^{1234}_{1110} + p^{1234}_{1101} + p^{1234}_{1111} \\
p^{14}_{00} &= p^{1234}_{0000} + p^{1234}_{0001} + p^{1234}_{1000} + p^{1234}_{1001} &
p^{14}_{10} &= p^{1234}_{0010} + p^{1234}_{0011} + p^{1234}_{1010} + p^{1234}_{1011} \\
p^{14}_{01} &= p^{1234}_{0100} + p^{1234}_{0101} + p^{1234}_{1100} + p^{1234}_{1101} &
p^{14}_{11} &= p^{1234}_{0110} + p^{1234}_{0111} + p^{1234}_{1110} + p^{1234}_{1111} \\
p^{32}_{00} &= p^{1234}_{0000} + p^{1234}_{0010} + p^{1234}_{0100} + p^{1234}_{0110} &
p^{32}_{10} &= p^{1234}_{1000} + p^{1234}_{1010} + p^{1234}_{1100} + p^{1234}_{1110} \\
p^{32}_{01} &= p^{1234}_{0001} + p^{1234}_{0011} + p^{1234}_{0101} + p^{1234}_{0111} &
p^{32}_{11} &= p^{1234}_{1001} + p^{1234}_{1011} + p^{1234}_{1101} + p^{1234}_{1111}\\
p^{34}_{00} &= p^{1234}_{0000} + p^{1234}_{1000} + p^{1234}_{0100} + p^{1234}_{1100} &
p^{34}_{10} &= p^{1234}_{0010} + p^{1234}_{1010} + p^{1234}_{0110} + p^{1234}_{1110} \\
p^{34}_{01} &= p^{1234}_{0001} + p^{1234}_{1001} + p^{1234}_{0101} + p^{1234}_{1101} &
p^{34}_{11} &= p^{1234}_{0011} + p^{1234}_{1011} + p^{1234}_{0111} + p^{1234}_{1111},
\end{aligned}
\end{equation}
which are displayed in matrix form in \autoref{tab:logmat222}.

Rather than proceeding to compute $\mathbb{L}(\mathcal{G})$ using elimination of inequalities as before, we will instead make use of the fact that the extremal points of $\mathbb{L}(\mathcal{G})$ happen to be the extremal points of $\mathbb{L}(\mathcal{G})$ with integer coordinates.  This is the approach which was used to compute the volume ratios shown in \autoref{fig:ncycvolrat}.  More specifically, those computations were done using a computer program based on the following algorithm which is available via a virtual machine that can be reconstructed using the instructions available on \href{https://github.com/cameronraysmith/sep}{github}:

\begin{enumerate}
\item Compute (a basis for) the cokernel of $\mathbf{G}$. The cokernel gives the obstructions to the system $\mathbf{G}\mathbf{X}=\mathbf{V}$ having a solution. In order to eliminate these obstructions constraints must be imposed on $\mathbb{R}^{\mathcal{E}(\mathcal{G})}$ and these constraints are given precisely via annihilating the cokernel.
\item Use the constraints on $\mathbb{R}^{\mathcal{E}(\mathcal{G})}$ from step 1 necessary for the system $\mathbf{G}\mathbf{X}=\mathbf{V}$ to have a solution to eliminate variables from the system of inequalities $\mathbf{V} \geq 0$ giving a half-space representation or H-representation of the polytope $\mathbb{L}(\mathcal{G})$. This can be used to compute $\text{Vol}(\mathbb{L}(\mathcal{G}))$.
\item Compute the vertices of $\mathbb{L}(\mathcal{G})$ from the H-representation determined in step 2 giving a vertex representation or V-representation of $\mathbb{L}(\mathcal{G})$.
\item Filter the non-integer rational vertices from the collection computed in step 3 to produce a corresponding V-representation of $\mathbb{M}(\mathcal{G})$ \cite{Wainwright2007} proposition 8.3.
\item Compute $\text{Vol}(\mathbb{M}(\mathcal{G}))$ from the V-represention of $\mathbb{M}(\mathcal{G})$.
\end{enumerate}

For standard computations on polytopes, we make use of the standard algorithms incorporated by the polymake project \cite{Gawrilow2000}. In some cases, the volume computation is too costly to perform exactly. In those cases we use the approximation given in \cite{Cousins}.  We now return to our example of four variables $\mathcal{G} = \{\{l_1,l_2 \},\{l_1,l_4 \},\{l_3,l_2\},\{l_3,l_4\} \}$ and $P=\{0,1\}$ and use it to walk through key components of the algorithm.

The equalities derived by computing the cokernel of the matrix $\mathbf{G}$ given in \autoref{tab:logmat222} and adjoining rows that enforce the normalization of the marginal distributions are represented as a matrix in \autoref{eq:eqmat}.
\begin{equation}\label{eq:eqmat}
\begin{aligned}
\begin{bmatrix}
  -1 & -1 & 0 & 0 & 1 & 1 & 0 & 0 & 0 & 0 & 0 & 0 & 0 & 0 & 0 & 0 & 0\\
  0 & 0 & -1 & -1 & 0 & 0 & 1 & 1 & 0 & 0 & 0 & 0 & 0 & 0 & 0 & 0 & 0\\
  -1 & 0 & -1 & 0 & 0 & 0 & 0 & 0 & 1 & 1 & 0 & 0 & 0 & 0 & 0 & 0 & 0\\
  0 & -1 & 0 & -1 & 0 & 0 & 0 & 0 & 0 & 0 & 1 & 1 & 0 & 0 & 0 & 0 & 0\\
  0 & 0 & 0 & 0 & 0 & 0 & 0 & 0 & -1 & 0 & -1 & 0 & 1 & 1 & 0 & 0 & 0\\
  0 & 0 & 0 & 0 & -1 & 0 & -1 & 0 & 0 & 0 & 0 & 0 & 1 & 0 & 1 & 0 & 0\\
  -1 & -1 & -1 & -1 & 1 & 0 & 1 & 0 & 1 & 0 & 1 & 0 & -1 & 0 & 0 & 1 & 0\\
  1 & 1 & 1 & 1 & 0 & 0 & 0 & 0 & 0 & 0 & 0 & 0 & 0 & 0 & 0 & 0 & 1\\
  0 & 0 & 0 & 0 & 1 & 1 & 1 & 1 & 0 & 0 & 0 & 0 & 0 & 0 & 0 & 0 & 1\\
  0 & 0 & 0 & 0 & 0 & 0 & 0 & 0 & 1 & 1 & 1 & 1 & 0 & 0 & 0 & 0 & 1\\
  0 & 0 & 0 & 0 & 0 & 0 & 0 & 0 & 0 & 0 & 0 & 0 & 1 & 1 & 1 & 1 & 1\\
\end{bmatrix}
\end{aligned}
\end{equation}
The final column represents the right-hand side of each equality. It turns out all but one of the normalization conditions is linearly dependent with respect to the other equalities and so we can reduce this set of $7+4=11$ constraints to the $8$ represented again in matrix form in \autoref{eq:kceqsrefa}.
\begin{equation}\label{eq:kceqsrefa}
\begin{aligned}
\begin{bmatrix}
  1 & 0 & 0 & -1 & 0 & 0 & 1 & 1 & 0 & 0 & 1 & 1 & 0 & 0 & 0 & 0 & 1\\
  0 & 1 & 0 & 1 & 0 & 0 & 0 & 0 & 0 & 0 & -1 & -1 & 0 & 0 & 0 & 0 & 0\\
  0 & 0 & 1 & 1 & 0 & 0 & -1 & -1 & 0 & 0 & 0 & 0 & 0 & 0 & 0 & 0 & 0\\
  0 & 0 & 0 & 0 & 1 & 0 & 1 & 0 & 0 & 0 & 0 & 0 & 0 & 1 & 0 & 1 & 1\\
  0 & 0 & 0 & 0 & 0 & 1 & 0 & 1 & 0 & 0 & 0 & 0 & 0 & -1 & 0 & -1 & 0\\
  0 & 0 & 0 & 0 & 0 & 0 & 0 & 0 & 1 & 0 & 1 & 0 & 0 & 0 & 1 & 1 & 1\\
  0 & 0 & 0 & 0 & 0 & 0 & 0 & 0 & 0 & 1 & 0 & 1 & 0 & 0 & -1 & -1 & 0\\
  0 & 0 & 0 & 0 & 0 & 0 & 0 & 0 & 0 & 0 & 0 & 0 & 1 & 1 & 1 & 1 & 1\\
\end{bmatrix}
\end{aligned}
\end{equation}
These equalities can now be substituted into the positivity inequalities necessary to define any space of probability distributions. This yields a set of inequalities \autoref{eq:polrepindineqs} that specify an H-representation of the polytope $\mathbb{L}(\mathcal{G})$. This is the modular polytope, which is a subspace of $\Delta_3^{\oplus 4}$ associated to distributions consistent with the linear transformation $\mathbf{G}$
\begin{equation}\label{eq:polrepindineqs}
\begin{aligned}
\begin{bmatrix}
  1 & 1 & -1 & -1 & -1 & -1 & 0 & 0 & 0\\
  0 & -1 & 0 & 0 & 1 & 1 & 0 & 0 & 0\\
  0 & -1 & 1 & 1 & 0 & 0 & 0 & 0 & 0\\
  1 & 0 & -1 & 0 & 0 & 0 & -1 & 0 & -1\\
  0 & 0 & 0 & -1 & 0 & 0 & 1 & 0 & 1\\
  1 & 0 & 0 & 0 & -1 & 0 & 0 & -1 & -1\\
  0 & 0 & 0 & 0 & 0 & -1 & 0 & 1 & 1\\
  1 & 0 & 0 & 0 & 0 & 0 & -1 & -1 & -1\\
  0 & 1 & 0 & 0 & 0 & 0 & 0 & 0 & 0\\
  0 & 0 & 1 & 0 & 0 & 0 & 0 & 0 & 0\\
  0 & 0 & 0 & 1 & 0 & 0 & 0 & 0 & 0\\
  0 & 0 & 0 & 0 & 1 & 0 & 0 & 0 & 0\\
  0 & 0 & 0 & 0 & 0 & 1 & 0 & 0 & 0\\
  0 & 0 & 0 & 0 & 0 & 0 & 1 & 0 & 0\\
  0 & 0 & 0 & 0 & 0 & 0 & 0 & 1 & 0\\
  0 & 0 & 0 & 0 & 0 & 0 & 0 & 0 & 1\\
\end{bmatrix}
\end{aligned}
\end{equation}
A row $(a_0,a_1,...,a_d)$ corresponds to the inequality $a_0 + a_1 x_1 + ... + a_d x_d >= 0$. The embedded identity matrix has, in this particular case eight, rows that specify the positivity of the variables corresponding to each of the, in this particular case eight, dimensions. Transforming this inequality or H-representation to a vertex or V-representation of the modular polytope produces \autoref{eq:vrepfromhrep}.
\begin{equation}\label{eq:vrepfromhrep}
\begin{aligned}
\begin{bmatrix}
  1 & 0 & 0 & 0 & 0 & 0 & 0 & 0 & 1\\
  1 & 0 & 0 & 0 & 0 & 0 & 0 & 1 & 0\\
  1 & 0 & 0 & 0 & 0 & 0 & 1 & 0 & 0\\
  1 & 1/2 & 1/2 & 0 & 1/2 & 0 & 1/2 & 1/2 & 0\\
  1 & 1/2 & 0 & 1/2 & 0 & 1/2 & 1/2 & 1/2 & 0\\
  1 & 0 & 0 & 0 & 0 & 0 & 0 & 0 & 0\\
  1 & 1/2 & 1/2 & 0 & 0 & 1/2 & 0 & 0 & 1/2\\
  1 & 1/2 & 0 & 1/2 & 1/2 & 0 & 0 & 0 & 1/2\\
  1 & 0 & 0 & 1/2 & 0 & 1/2 & 0 & 0 & 1/2\\
  1 & 0 & 1/2 & 0 & 1/2 & 0 & 0 & 0 & 1/2\\
  1 & 0 & 1/2 & 0 & 0 & 1/2 & 1/2 & 1/2 & 0\\
  1 & 0 & 0 & 1/2 & 1/2 & 0 & 1/2 & 1/2 & 0\\
  1 & 1 & 1 & 0 & 1 & 0 & 0 & 0 & 0\\
  1 & 0 & 0 & 0 & 1 & 0 & 0 & 0 & 0\\
  1 & 0 & 1 & 0 & 0 & 0 & 0 & 0 & 0\\
  1 & 0 & 0 & 1 & 0 & 0 & 1 & 0 & 0\\
  1 & 0 & 0 & 0 & 0 & 1 & 0 & 1 & 0\\
  1 & 0 & 0 & 0 & 1 & 0 & 1 & 0 & 0\\
  1 & 0 & 1 & 0 & 0 & 0 & 0 & 1 & 0\\
  1 & 1 & 0 & 1 & 0 & 1 & 0 & 0 & 1\\
  1 & 1 & 0 & 1 & 1 & 0 & 1 & 0 & 0\\
  1 & 1 & 1 & 0 & 0 & 1 & 0 & 1 & 0\\
  1 & 0 & 0 & 1 & 0 & 0 & 0 & 0 & 1\\
  1 & 0 & 0 & 0 & 0 & 1 & 0 & 0 & 1\\
\end{bmatrix}
\end{aligned}
\end{equation}
This completes steps 1-3 of the algorithm outlined above. Step 4 is trivial; to obtain the V-represention of $\mathbb{M}(\mathcal{G})$, we strike out the rows in which $1/2$ appears.  Finally, we compute the volume of the polytope whose vertices are the rows of \autoref{eq:vrepfromhrep} to obtain $\text{Vol}(\mathbb{L}(\mathcal{G})) = \frac{1}{120}$ and the volume of the polytope whose vertices are rows of integers to obtain $\text{Vol}(\mathbb{M}(\mathcal{G})) = \frac{1}{180}$ yielding a ratio $\frac{\text{Vol}(\mathbb{M}(\mathcal{G}))}{\text{Vol}(\mathbb{L}(\mathcal{G}))} = \frac{2}{3}$.

\FloatBarrier
\pagebreak

\pagebreak

\pagebreak

\pagebreak

\pagebreak

\FloatBarrier
\pagebreak

\begin{table}[!ht]
\centering
\begin{tabular}{ r || c | c | c | c | c | c | c | c | c | c | c | c | c | c | c | c }
		 &	\begin{sideways}$e^{1234}_{0000}$\end{sideways} & \begin{sideways}$e^{1234}_{0010}$\end{sideways} & \begin{sideways}$e^{1234}_{0001}$\end{sideways} & \begin{sideways}$e^{1234}_{0011}$\end{sideways}
			  & \begin{sideways}$e^{1234}_{1000}$\end{sideways} & \begin{sideways}$e^{1234}_{1010}$\end{sideways} & \begin{sideways}$e^{1234}_{1001}$\end{sideways} & \begin{sideways}$e^{1234}_{1011}$\end{sideways}
			  &	\begin{sideways}$e^{1234}_{0100}$\end{sideways} & \begin{sideways}$e^{1234}_{0110}$\end{sideways} & \begin{sideways}$e^{1234}_{0101}$\end{sideways} & \begin{sideways}$e^{1234}_{0111}$\end{sideways}
			  &	\begin{sideways}$e^{1234}_{1100}$\end{sideways} & \begin{sideways}$e^{1234}_{1110}$\end{sideways} & \begin{sideways}$e^{1234}_{1101}$\end{sideways} & \begin{sideways}$e^{1234}_{1111}$\end{sideways}\\ \hline \hline
    $e^{12}_{00}$ & 1 & 1 & 1 & 1 & 0 & 0 & 0 & 0 & 0 & 0 & 0 & 0 & 0 & 0 & 0 & 0\\ \hline
    $e^{12}_{10}$ & 0 & 0 & 0 & 0 & 1 & 1 & 1 & 1 & 0 & 0 & 0 & 0 & 0 & 0 & 0 & 0\\ \hline
    $e^{12}_{01}$ & 0 & 0 & 0 & 0 & 0 & 0 & 0 & 0 & 1 & 1 & 1 & 1 &  0 & 0 & 0 & 0\\ \hline
    $e^{12}_{11}$ & 0 & 0 & 0 & 0 & 0 & 0 & 0 & 0 & 0 & 0 & 0 & 0 & 1 & 1 & 1 & 1\\ \hline

    $e^{32}_{00}$ & 1 & 0 & 1 & 0 & 1 & 0 & 1 & 0 & 0 & 0 & 0 & 0 & 0 & 0 & 0 & 0\\ \hline
    $e^{32}_{10}$ & 0 & 1 & 0 & 1 & 0 & 1 & 0 & 1 & 0 & 0 & 0 & 0 & 0 & 0 & 0 & 0\\ \hline
    $e^{32}_{01}$ & 0 & 0 & 0 & 0 & 0 & 0 & 0 & 0 & 1 & 0 & 1 & 0 & 1 & 0 & 1 & 0\\ \hline
    $e^{32}_{11}$ & 0 & 0 & 0 & 0 & 0 & 0 & 0 & 0 & 0 & 1 & 0 & 1 & 0 & 1 & 0 & 1\\ \hline

    $e^{14}_{00}$ & 1 & 1 & 0 & 0 & 0 & 0 & 0 & 0 & 1 & 1 & 0 & 0 & 0 & 0 & 0 & 0\\ \hline
    $e^{14}_{10}$ & 0 & 0 & 0 & 0 & 1 & 1 & 0 & 0 & 0 & 0 & 0 & 0 & 1 & 1 & 0 & 0\\ \hline
    $e^{14}_{01}$ & 0 & 0 & 1 & 1 & 0 & 0 & 0 & 0 & 0 & 0 & 1 & 1 & 0 & 0 & 0 & 0\\ \hline
    $e^{14}_{11}$ & 0 & 0 & 0 & 0 & 0 & 0 & 1 & 1 & 0 & 0 & 0 & 0 & 0 & 0 & 1 & 1\\ \hline

    $e^{34}_{00}$ & 1 & 0 & 0 & 0 & 1 & 0 & 0 & 0 & 1 & 0 & 0 & 0 & 1 & 0 & 0 & 0\\ \hline
    $e^{34}_{10}$ & 0 & 1 & 0 & 0 & 0 & 1 & 0 & 0 & 0 & 1 & 0 & 0 & 0 & 1 & 0 & 0\\ \hline
    $e^{34}_{01}$ & 0 & 0 & 1 & 0 & 0 & 0 & 1 & 0 & 0 & 0 & 1 & 0 & 0 & 0 & 1 & 0\\ \hline
    $e^{34}_{11}$ & 0 & 0 & 0 & 1 & 0 & 0 & 0 & 1 & 0 & 0 & 0 & 1 & 0 & 0 & 0 & 1\\
    \end{tabular}
\caption{Explicit construction of $\mathbf{G}_{n \times m}$ for the case $L = \{ l_1,l_2,l_3,l_4 \}$, $\mathcal{G} = \{\{l_1,l_2 \},\{l_1,l_4 \},\{l_3,l_2\},\{l_3,l_4\} \}$, $P=\{0,1\}$ and thus $\mathbf{G}_{(2 \cdot 2)^2 \times 2^{2 \cdot 2}} = \mathbf{G}_{16 \times 16}$.}
\label{tab:logmat222}
\end{table}

\end{document}